\documentclass[journal,10pt,twocolumn]{IEEEtran}
\IEEEoverridecommandlockouts
\usepackage[utf8]{inputenc}
\usepackage{amsmath}
\usepackage{amssymb}
\usepackage{mathtools}
\usepackage{cite}
\usepackage{amsmath,amssymb,amsfonts}
\usepackage{algorithmic}
\usepackage{graphicx}
\usepackage{textcomp}
\usepackage{textcomp} 
\usepackage{xcolor}
\usepackage{algorithmic}
\usepackage{algorithm}
\usepackage{hyperref}

\usepackage{subcaption,caption}
\usepackage{multirow}
\usepackage{tabularx}
\usepackage{flushend}

\usepackage[top    = 0.750in,
bottom = 1.05in,
left   = 0.75in,
right  = 0.75in]{geometry}
\def\BibTeX{{\rm B\kern-.05em{\sc i\kern-.025em b}\kern-.08em
    T\kern-.1667em\lower.7ex\hbox{E}\kern-.125emX}}

\begin{document}

\flushend
\title{Multi-Objective Optimization for Joint Communication and Sensing in Multi-user MIMO Systems: Characterizing the Pareto Boundary

%Characterization of Pareto Boundary for Joint Communication and Sensing in Multi-user MIMO Wireless Systems

%\thanks{The work was supported by the Department of National Defence (DND), Canada, through  
%an Innovation for Defence Excellence and Security (IDEaS) - Innovation Networks:
%Micro-nets 
%project.}
}

\author{\IEEEauthorblockN{Thakshila Perera, \textit{Graduate Student Member, IEEE}, Amine Mezghani,\textit{Member, IEEE}, \\ and Ekram Hossain,  \textit{Fellow, IEEE} }
\vspace{-10 mm}
\thanks{The authors are with the Department of Electrical and Computer Engineering at the University of Manitoba, Winnipeg, MB R3T 2N2, Canada (e-mail: pererat1@myumanitoba.ca; Amine.Mezghani@umanitoba.ca; Ekram.Hossain@umanitoba.ca). 

A part of the paper was presented in IEEE Int. Conf. on Communications 2024 (ICC'24)~\cite{10615441}. 

The work was supported by the Department of National Defence (DND), Canada, through  
an Innovation for Defence Excellence and Security (IDEaS) - Innovation Networks:
Micro-nets project.}
}

\maketitle
\begin{abstract}
This paper investigates the Pareto boundary performance of a joint communication and sensing (JCAS) system that addresses both sensing and communication functions at the same time. In this scenario, a multiple-antenna base station (BS) transmits information to multiple single-antenna communication users while concurrently estimating the parameters of a single sensing object using the echo signal. We present an integrated beamforming approach for JCAS in a multi-user  multiple-input and multiple-output (MIMO) system. The performance measures for communication and sensing are Fisher information (FI) and mutual information (MI). Our research considers two scenarios: multiple communication users with a single sensing object and a single communication user with a single sensing object. We formulate a multi-objective optimization problem to maximize the weighted sum of MI and FI, subject to a total transmit power budget for both cases. As a particular case, we address the equivalent isotropic radiated power (EIRP) for the single communication user scenario. We use the uplink-downlink duality for the multi-user case to simplify the problem and apply Lagrangian optimization and line search methods with a block-coordinate ascending technique. We use projected gradient descent (PGD) to solve the optimization problem in the single-user case. Our numerical results demonstrate that joint beamforming is optimal for the multi-user JCAS system, as opposed to independent beamforming for each user and the sensing object. Furthermore, we reveal the Pareto boundary for the multi-user case, with variations in the number of communication users and the number of transmitting and receiving antennas. We provide the Pareto boundary depending on EIRP limitations for the single-user case.
\end{abstract}


\section{Introduction}
\subsection{Background and Related Works}
Spurred by the rising demand for advanced services and the explosion of IoT devices, the next generation (6G and beyond) systems seek to transform wireless communication networks. These future networks will be crucial for emerging applications such as intelligent transportation and smart homes, requiring superior wireless connectivity and strong sensing capabilities. The vision is for these networks to integrate communication and sensing capabilities, offering extensive connectivity and meeting the growing demands for spectrum and quality of service\cite{b2d30f91f67f4698946766bd16786c74,10008554,9737357}.

Integrated Sensing and Communication (ISAC) is a fundamental concept that combines sensing technologies (such as radar or sensors) with communication capabilities within a single system. This integration allows sharing the same frequency bands and hardware, thereby reducing costs \cite{9705498,9606831,10418473}. Joint Communication and Sensing (JCAS) is a specific application within ISAC that focuses on radar functionalities \cite{9705498}. JCAS can be categorized based on the integration of communication and sensing into two types: (i) the combination of both functionalities within a single system employing distinct hardware components and different frequency waveforms or timing, and (ii) the integration of both functions via shared hardware components, utilizing a singular waveform optimized to augment both communication and sensing capabilities \cite{9540344}. Contemporary JCAS systems can be classified into three categories: (i) communication-centric design, (ii) radar-centric design, and (iii) joint design and optimization \cite{9963506}.

Integrating JCAS into wireless networks enhances spectrum efficiency, coverage, and signal quality by dynamically allocating resources and optimizing beamforming based on real-time environmental data \cite{8905229,9843013, pucci2022systemlevel}. 
Additionally, it improves energy efficiency by intelligently scheduling communication and sensing tasks, enhancing network security and privacy through real-time threat detection and anomaly monitoring. Furthermore, JCAS enables innovative applications like autonomous navigation, augmented reality, and smart city solutions, driving consumer adoption and maximizing the potential of 5G technology \cite{9354629}.

Integrating multiple antennas at both the transmitting and receiving ends (known as MIMO) has become essential for implementing modern and future wireless communication systems  \cite{hashem2015critical}. MIMO is integrated into recent communication standards such as IEEE 802.11, Long Term Evolution (LTE), and 5G \cite{sibille2010mimo,6798744}. Furthermore, MIMO signaling enhances communication systems through techniques such as spatial multiplexing and diversity methods \cite{1341263,7034415}. MIMO radar systems were inspired by MIMO communication, incorporating angular diversity to improve radar performance \cite{1399142}.

MIMO JCAS is crucial for enhancing both communication and sensing performances. In particular, researchers focus on MIMO JCAS because it presents a key challenge in tackling optimization problems to design beamforming based on communication and sensing performance metrics \cite{10251151}. In the JCAS system, MI (or sum-rate capacity) and FI (or Cramér-Rao Lower Bound (CRLB)) can be used as fundamental performance metrics for communication and sensing, respectively \cite{li2023framework,9764187, Liu_2022}. The trade-off between effective communication and accurate sensing will be a critical factor to consider in real-world applications.

The FI matrix quantifies the amount of information an observable random variable contains about unknown parameters in the statistical model. The CRLB is a theoretical lower bound on the variance of any unbiased parameter estimator, providing a lower limit on how precise an impartial estimator can be \cite{kay1993fundamentals,liu2021surveyfundamentallimitsintegrated}. The authors in \cite{Liu_2022,9764187} studied MIMO joint radar and communication systems, designing the precoding based on minimizing the CRB with signal-to-interference-plus-noise ratio (SINR) for single and multiple users with point and extended target cases. Furthermore, the CRB-rate trade-off comes to the forefront instead of CRB optimization. \cite{10217169} and \cite{xiong2023fundamentaltradeoffintegratedsensing} investigate the Pareto boundary between communication and sensing within the CRB-rate region. However, \cite{10217169} considered the MIMO ISAC model, while \cite{xiong2023fundamentaltradeoffintegratedsensing} focused on the point-to-point (P2P) ISAC model.

MI  is a crucial factor in ensuring the quality of service (QoS) for data transmission from the BS to communication users \cite{li2023framework}. In systems with multiple users and channels, the sum-rate capacity concept represents the highest achievable total data rate across all users or channels \cite{HeathJr._Lozano_2018}. In\cite{liu2023deterministicrandomtradeoffintegratedsensing}, the authors studied the deterministic-random trade-off between communication and sensing in a generic P2P downlink ISAC system with one or multiple targets using a Gaussian ISAC channel model. They employed conditional MI as a performance metric for both communication and sensing. Additionally, \cite{10129042} considered an uplink and downlink ISAC system over Gaussian channels, focusing on MI as a metric for both communication and sensing to reveal the trade-off between these functions. In addition, \cite{10038611} studied a downlink MIMO ISAC system with frequency-division sensing and communication (FDSAC). This system involved multiple users and targets to establish the Pareto boundary between communication and sensing. They used MI as the performance metric for both functions. Moreover, \cite{9800940} conducted a comprehensive study of an uplink and downlink ISAC system that involved multiple users and targets. The goal was to find the Pareto boundary between communication and sensing, using MI as the performance metric for both functions.

From the perspective of information theory, using a multiple-access channel (MAC) makes it easier to deal with the broadcast channel (BC) in MIMO multi-user systems due to differences in interference and cooperation \cite{4411629}. Researchers have used uplink-downlink duality in MIMO-related studies to address the downlink problem \cite{miretti2024uldldualitycellfreemassive,7510762,5413251}. In \cite{attiah2024beamformingdesignintegratedsensing}, the authors considered a multi-user ISAC system with $K$ single-antenna users. They minimized the Bayesian CRB (B-CRB) of the sensing problem while satisfying the quality of service constraints for the communication users. Moreover, they considered the uplink-downlink duality relation to simplify the model.

Nevertheless, all MIMO JCAS or ISAC systems consider MI or CRB-based optimization problems when designing beamforming while achieving the trade-off between communication and sensing. With these considerations, we are motivated to find the Pareto boundary for communication and sensing using MI as the communication performance metric and FI as the sensing performance metric. Moreover, the uplink-downlink duality relationship in MI and FI in MIMO JCAS systems is another motivating factor for our research.
\vspace{-10pt}
\subsection{Contribution}
This paper proposes a method to jointly optimize beamforming for communication and sensing in a MIMO JCAS system. We consider two cases: multi-user communication with a single sensing object and single-user communication with a single sensing object. In this method, we treat all types of interference as noise and assume no interference between users and the sensing object. Instead of using MI or CRB as the performance metrics for both communication and sensing, we use MI and FI separately as communication and sensing performance metrics, respectively. Moreover, we focus mainly on the Pareto boundary between communication and sensing regarding FI and MI. For the single-user case, we consider the EIRP constraint at the BS to limit the spatial focusing of the signal. With this framework, the Pareto boundary will serve as a key benchmark in actual implementation due to the consideration of both FI and MI. In summary,
\begin{itemize}
    \item We formulate a multi-objective optimization problem to maximize the weighted sum of MI and FI under the total transmit power budget for both cases and consider EIRP limitation as a constraint in the single-user case.

    \item For the multi-user communication case, we use the uplink-downlink duality concept to simplify the primary objective function, and due to the convexity of the converted problem, we use the Lagrangian optimization technique and line search method to solve the problem. Moreover, we consider the block coordinate ascending technique enabled through the MAC to BC conversion.
    
    \item For the single-user communication case, the optimization problem is convex. However, with the new constraint of EIRP, we use the PGD technique with the Karush-Kuhn-Tucker (KKT) method to solve the problem.

    \item In simulation results, we demonstrate the Pareto boundary with MI and FI for both cases. Moreover, we illustrate the dependencies of the communication and sensing performances w.r.t. the number of antennas at the BS, number of downlink users of the system, signal-to-noise (SNR) ratio of the BS, and EIRP limitation at the BS.
\end{itemize}
\subsection{Paper Organization and Notations}
\vspace{0.1cm}
\noindent
\textbf{Organization}: We organize the remainder of this paper as follows: Section II presents the JCAS-enabled MIMO system model and assumptions, along with the formulation of MI and FI based on MAC to BC conversion. Section III discusses the multi-objective optimization problem with the corresponding constraints for the multi-user communication case and the corresponding solution approach. Section IV proposes the multi-objective maximization problem with the constraints and the solution approach for the single-user communication case. Simulation results are provided in Section V. Finally, Section VI concludes this paper.

\vspace{0.1cm}
\noindent
\textbf{Notations}: Vectors and matrices are denoted by small boldface letters (e.g., $\mathbf{v}$) and capital boldface letters (e.g., $\mathbf{X}$), respectively. $\mathbb{C}^{N \times M}$ denotes the space of $N \times M$ complex matrices. $\mathbf{I}$ represents the identity matrix with appropriate dimensions. For a square matrix $\mathbf{A}$, $\operatorname{tr}(\mathbf{A})$ denotes its trace , $\mathbf{A} \succeq 0$ means that $\mathbf{A}$ is positive semi-definite and $\mathbf{A} \preceq 0$ denotes that $\mathbf{A}$ is negative semi-definite. For any arbitrary complex matrix $\mathbf{B}$, $\mathbf{B}(i,j)$, $\text{rank}(\mathbf{B})$, $\mathbf{B}^T$, $\mathbf{B}^H$, and $\mathbf{B}^*$ denotes its $(i,j)^{th}$ element, rank, transpose, conjugate transpose, and complex conjugate respectively. $\mathbb{E}(\cdot)$ denotes the statistical expectation. $||\cdot||$ is the Euclidean norm of a vector, $||\cdot||_\mathbf{F}$ is the Frobenius norm of a matrix, and $|\cdot|$ is the absolute value. 
%\vspace{-1mm}
\section{System Model and Assumptions}
We consider a JCAS-enabled MIMO BS equipped with two separate antenna arrays: $N_t$	transmitting antennas and $N_r$ receiving antennas. As shown in Fig.~\ref{Fig1}, the MIMO BC system model includes $K$ single-antenna downlink users (where $K < N_t$) and a sensing object. We assume that line-of-sight (LOS) and non-line-of-sight (NLOS) conditions affect downlink communication, while only LOS affects sensing. Also, we assume that there is no interference between each downlink user and between the users and the sensing object. Furthermore, all other interferences and noises are considered additive white Gaussian noise (AWGN).

\subsection{Signal Model for Multi-user Communication Case}
\begin{figure}[!t]
\centering\vspace{0mm}
\includegraphics[scale = 0.4]{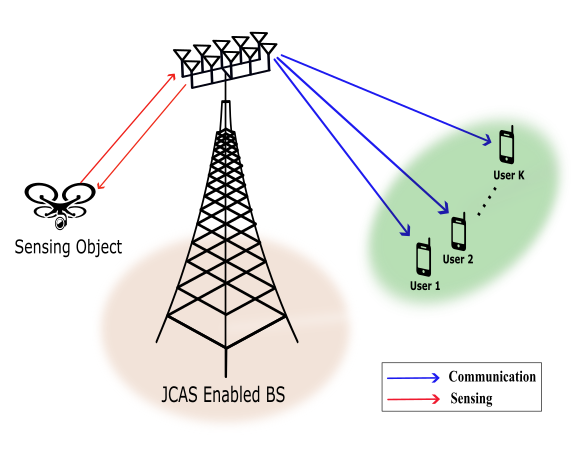}
\caption{Joint communication and sensing model of MIMO BC.}
\label{Fig1}
\vspace{-5mm}
\end{figure}

As shown in the MIMO BC system model (Fig.~\ref{Fig1}), we represent the JCAS signal from the BS as $\mathbf{x}_{{BC}} \in \mathbb{C}^{N_t}$:
\begin{equation}
\label{eqn:e3}
\mathbf{R}_{x} = {\mathbb{E}}[\mathbf{x}_{{BC}} \mathbf{x}_{{BC}}^H] = \sum_{i=1}^{K} \mathbf{w}_{i}\mathbf{w}_i^H,
\end{equation}
where $\mathbf{w}_{i} \in \mathbb{C}^{N_t}$ is the beamforming vector for the $i^\text{th}$ downlink user.
We define the JCAS downlink covariance matrices $\mathbf{R}_{i}  \in \mathbb{C}^{N_t \times N_t} $ for $i$ = 1,...,$K$ with the total power constraint $\sum_{i=1}^{K} \operatorname{tr}(\mathbf{R}_{i}) \leq  P_{tx} $, where $P_{tx}$ is the transmit power budget. The received signal by the communication users of  the MIMO BC is
\begin{equation}
\label{eqn:MIMOBC}
    {y}_{i} = \mathbf h_i^H \mathbf{x}_{{BC}} + {z}_{i},
\end{equation}
in which $\mathbf h_i^H \in \mathbb{C}^{1 \times N_t}$ is the downlink channel vector of the $i^{th}$ user. The random variable ${z}_{i}$ is the AWGN  with zero mean and $\sigma^2_{c}$ variance.

Let us consider the MIMO MAC system model for the user side of Fig.~\ref{Fig1}. The JCAS signal to the BS from the uplink users is represented as $x_{i} \in \mathbb{C}$. We define the uplink covariance matrices ${{Q}}_{i}  \in \mathbb{C} $ with the total power constraint of $\sum_{i=1}^{K} \operatorname{tr}({Q}_{i}) \leq  P_{tx} $. 
%(The covariance matrix becomes a 1x1 due to each user having a single antenna.)
The received signal at the BS from users of the MIMO MAC is
\begin{equation}
\label{eqn:MIMOMAC}
    \mathbf{y}_{{MAC}} = \sum_{i=1}^{K} \mathbf h_i {x}_{i} + \mathbf{z},
\end{equation}
where $\mathbf h_i$ is the dual uplink channel vector for the $i^{th}$ user and $\mathbf{z}$ is the AWGN with zero mean and $\sigma^2_{c}$ variance for all elements.

On the sensing side, the BS transmits $\mathbf{x}_{{BC}}$ to the sensing object, and the BS receives the reflected echo signal 
\begin{equation}
\label{eqn:sensing}
    \mathbf {y}_{r} = \mathbf{G} \mathbf{x}_{{BC}} + \mathbf{z}_{r},
\end{equation}
where $\mathbf z_{r}\in \mathbb{C}^{N_r}$ is the AWGN with zero mean and $\sigma^2_{r}$ variance for each. $\mathbf{G}\in \mathbb{C}^{N_r \times N_t}$ denotes the target response matrix, and for a single target case, it can be represented as:
\begin{equation}
\label{eqn:G}
    \mathbf G = \gamma_{r} \mathbf{b}{(\theta_{r})}\mathbf a^{T}{(\theta_{r})}, \; \mbox{with}
\end{equation}
\begin{align}
\label{eqn:atheta}
    {\mathbf {a}}\left(\theta_r \right) = {\left[ {{e^{ - j\frac{{{N_t-1}}}{2} k \sin \theta_r }},{e^{ - j\frac{{{N_t} - 3}}{2} k \sin \theta_r }}, \ldots, {e^{j\frac{{{N_t-1}}}{2} k \sin \theta_r }}} \right]^T}, 
\end{align} 
where $\gamma_r \in \mathbb{C}$ is the reflection coefficient which includes both radar cross-section and round trip path-loss, and $\theta_r$ is the angle of the sensing object relative to the BS and where $k= \frac{2\pi d}{\lambda}$, $d$ is the distance between two adjacent antennas in the array,  $\lambda$ is the wavelength. The vector $\mathbf a{(\theta_{r})} \in \mathbb{C}^{N_t}$ (as given in \eqref{eqn:atheta}) denotes the transmit array response vector assuming that the antenna array has its origin located at the center. Similarly, we can express  $\mathbf{b}{(\theta_{r})} \in \mathbb{C}^{N_r}$, which is the receive array response vector. In our case, the angles of arrival (AoA) and departure (AoD) are equal (i.e., a monostatic radar).
% Figure 1
%%===================================================================
%%===================================================================

% Figure 1
%%===================================================================
%\begin{figure}[!t]
%\centering\vspace{0mm}
%\includegraphics[scale = 0.4]%{system_model_multiuser_MIMO_MAC.png}
%\caption{Joint communication and sensing system model of MIMO MAC.}
%\label{Fig2}
%\vspace{-5mm}
%\end{figure}
%%===================================================================

\subsection{Mutual Information for the Communication User}

In \cite{1412050}, \cite{liu2007conjugate}, \cite{1327794} and \cite{1207369}, the sum-rate capacity of the MIMO BC (denoted as ${C}_{BC}$) can be achieved by the dirty paper coding \cite{1056659}. Then, the sum-rate capacity of the MIMO BC can be expressed  in terms of maximization with the power constraint $\sum_{i=1}^{K} \operatorname{tr}(\mathbf{R}_{i}) \leq  P_{tx}$ as follows:
\begin{align} \label{MIMOBCCAP}
    {C}_{BC}  & = \max\limits_{\{\mathbf{R}_{i}\}_{i=1}^K }  \sum_{i=1}^{K} \log{\left \vert 1+\mathbf{h}_{i}^H({\sum}_{j=1}^{i}\mathbf{R}_{j})\mathbf{h}_{i}\right \vert \over \left \vert 1+\mathbf{h}_{i}^H({\sum}_{j=1}^{i-1}\mathbf{R}_{j}) \mathbf{h}_{i}\right \vert }.
\end{align}

We observe that \eqref{MIMOBCCAP} is not a concave function. Therefore, the optimization problem concerning $\mathbf{R}_i$ is nontrivial.
However, with the duality of the uplink and downlink established in the dirty paper region, the MIMO capacity, $C_{BC}$ is equal to the MIMO MAC capacity (${C}_{MAC}$)\cite{1412050}, \cite{1237143}:
\begin{equation} \label{BCMACEQUAL}
   {C}_{BC} (\mathbf{h}_{1}^{H}, ... , \mathbf{h}_{K}^{H},P_{tx}) = {C}_{MAC} (\mathbf{h}_{1}, ...,\mathbf{h}_{K},P_{tx}).
\end{equation}
The sum-rate capacity of the MIMO MAC with the power constraint of $\sum_{i=1}^{K} \operatorname{tr}({Q}_{i}) \leq  P_{tx}$ is 
\begin{equation}\label{MIMOMACCAP}  
    {C}_{MAC} = \max \limits_{\{{Q}_{i}\}_{i=1}^K} \log_2{ \left \vert \mathbf{I} + \sum_{i=1}^{K} \mathbf{h}_{i}{Q}_{i}\mathbf{h}_{i}^H\right \vert }.
\end{equation}

We can rewrite the ${C}_{MAC}$ as follows:
\begin{align}\label{MACP2P}  
    {C}_{MAC} &= \log_2{ \left \vert \mathbf{I} + \sum_{j \neq 1}^{K} \mathbf{h}_{j} {Q}_{j}\mathbf{h}_{j}^H + \mathbf{h}_\mathbf{i} {Q}_{i}\mathbf{h}_{i}^H  \right \vert} 
    \\ \nonumber 
    &= \log_2{\left \vert  \mathbf{A}_{M}^{1/2} \left(\mathbf{I}+\mathbf{A}_{M}^{-1/2}\mathbf{B}_{M}\mathbf{A}_{M}^{-1/2}\right)\mathbf{A}_{M}^{1/2}\right \vert} 
    \\ \nonumber
    &=  \log_2{\left \vert \mathbf{A}_{M}\left(\mathbf{I}+\mathbf{A}_{M}^{-1/2}\mathbf{B}_{M}\mathbf{A}_{M}^{-1/2}\right) \right \vert} 
    \\ \nonumber
    &= \log_2{\left \vert \mathbf{A}_{M}\right \vert} + \log_2{\left \vert \mathbf{I}+\mathbf{A}_{M}^{-1/2}\mathbf{B}_{M}\mathbf{A}_{M}^{-1/2} \right \vert},
\end{align}
where $\mathbf{A}_{M} = \mathbf{I} + \sum_{j \neq 1}^{K} \mathbf{h}_{j} {Q}_{j}\mathbf{h}_{j}^H $ and $\mathbf{B}_{M} = \mathbf{h}_{i} {Q}_{i}\mathbf{h}_{i}^H $.  \eqref{MACP2P} is equivalent to the capacity of point-to-point MIMO channel with effective channel of $\mathbf{h}_{ei}^H = \mathbf{h}_{i}^H\mathbf{A}_{M}^{-1/2}$ with the power constraint of $\operatorname{tr}({Q}_{i}) \leq  P_{i}$ \cite{1412050}.
\begin{equation}\label{MACCAPFINAL}  
    {C}_{MAC} = \max \limits_{\{{Q}_{i}\}_{i=1}^K: \sum_{i=1}^{K} \operatorname{tr}({Q}_{i}) \leq  P_{tx}} \sum_{i=1}^{K}  \log_2{ \left \vert   \mathbf{I} + \mathbf{h}_{ei} {Q}_{i}\mathbf{h}_{ei}^H \right \vert }.
\end{equation}

Treating these $\mathbf{h}_{ei}$ effective channels as parallel and non-interfering, we can obtain the sum-rate capacity as in \eqref{MACCAPFINAL}.

\subsection{Fisher Information for Sensing Object}
With the observation model in \eqref{eqn:sensing}, the BS knows $\mathbf{x}_{BC}$, and using the estimating vector $\mathbf{v}=[\theta_r \quad \gamma_r \quad \gamma_r^{*}]^T$, we can express the FI matrix, $\mathbf{J}_i$, along with each user (see $\textbf{Appendix  A}$):

\begin{align}
\label{eqn:FIM}
    \mathbf{J}_{i} =\frac{1}{\sigma^2_r} \begin{bmatrix}
        J_{(1,1)}
        &  J_{(1,2)} & J_{(1,3)}\\  J_{2,1)}& J_{(2,2)} & 0 \\J_{(3,1)} & 0 & J_{(3,3)}, 
    \end{bmatrix},   \; \mbox{with}
\end{align}
% with 
\begin{equation}
\label{eqn:FIJ1}
    J_{(1,1)} = 2\gamma_r^*\gamma_r \operatorname{tr}(\dot{\mathbf{A}} \mathbf{R}_{{i}} \dot{\mathbf{A}}^H )
\end{equation} 
\begin{equation}
\label{eqn:FIJ2}
    J_{(2,2)} = \operatorname{tr}(\mathbf{A} \mathbf{R}_{{i}} \mathbf{A}^H) 
\end{equation}
\begin{equation}
\label{eqn:FIJ3}
    J_{(3,3)} = \operatorname{tr}(\mathbf{A} \mathbf{R}_{{i}} \mathbf{A}^H) 
\end{equation}
\begin{equation}
\label{eqn:FIJ4}
    J_{(1,2)} = J_{(2,1)} = \gamma_r^*\operatorname{tr}(\mathbf{A} \mathbf{R}_{{i}} \dot{\mathbf{A}}^H)
\end{equation} 
\begin{equation}
\label{eqn:FIJ5}
    J_{(1,3)} = J_{(3,1)} = \gamma_r\operatorname{tr}(\dot{\mathbf{A}} \mathbf{R}_{{i}} \mathbf{A}^H),
\end{equation} 
where $\dot{\mathbf{A}} =\frac{\partial \mathbf{A}}{\partial \theta_r}  $ and $\mathbf{A} = \mathbf{b}{(\theta_{r})}\mathbf a^{T}{(\theta_{r})} $. The derivative of the transmit array response vector is mentioned in \eqref{eqn:athetadot}; similarly, we can write the derivative of the receive array response vector as 
\begin{align}
\label{eqn:athetadot}
    {\dot{\mathbf {a}}}\left(\theta_r \right) = {\left[ { - j{a_1}\frac{{{N_t-1}}}{2} k \cos \theta_r, \ldots, -j{a_{{N_t}}}\frac{{{N_t-1}}}{2} k \cos \theta_r } \right]^T}.
\end{align}

Regarding the sensing performance metric, we use the trace of the FI matrix due to its mathematical tractability. We are motivated by our observation that the parameters $\gamma_r$ and $\theta_r$ do not strongly couple, and the FI matrix remains mainly diagonal. Therefore, we consider the trace of $\mathbf{J}$ as follows:
\begin{equation}\label{traceJ}
    \operatorname{tr}(\mathbf{J})=\frac{2}{\sigma^2_r}  (\gamma_r^*\gamma_r \operatorname{tr}(\dot{\mathbf{A}} \mathbf{R}_{{x}} \dot{\mathbf{A}}^H )+\operatorname{tr}(\mathbf{A} \mathbf{R}_{{x}} \mathbf{A}^H )),
\end{equation}
where $\mathbf{R}_{{x}} = \sum_{i=1}^{K}\mathbf{w}_{i}\mathbf{w}_{i}^{H} + \mathbf{w}_0\mathbf{w}_0^H= \sum_{i=1}^{K} \mathbf{R}_{i}+\mathbf{R}_0$ with the $\mathbf{w}_0$ is the beamforming vector for the sensing object. Using the properties of trace, we can simplify \eqref{traceJ} as follows:
\begin{align}\label{traceJSimp}
     \operatorname{tr}(\mathbf{J}) &= \operatorname{tr}\left(\mathbf{M} \left(\sum_{i=1}^K\mathbf{R}_{i}+\mathbf{R}_0\right)\right),
     %\\ \nonumber &= \sum_{i=1}^K \operatorname{tr}(\mathbf{M}\mathbf{R}_{i}),
\end{align}
where $\mathbf{M} =  \frac{2\gamma_r^*\gamma_r}{\sigma^2_{r}} \dot{\mathbf{A}}^H \dot{\mathbf{A}} + \frac{2}{\sigma^2_{r}}\mathbf{A}^H \mathbf{A}$.

%\begin{align}\label{BCtraceFIM}
    %\operatorname{tr}(\mathbf{J}) &= \sum_{i=1}^K \operatorname{tr}(\mathbf{M}\mathbf{c}_{i} {Q}_{i}\mathbf{c}_{i}^H).
%\end{align}

%According to the uplink-downlink duality, we can use the MAC to BC transformation (see: \textbf{Appendix B}) to transform the downlink beamforming ($\mathbf{R}_\mathbf{i}$) matrix to uplink beamforming (${Q}_{i}$) due to the purpose of further problem formulations. With the MAC to BC transformation, the trace of the FI matrix can be expressed as in \eqref{BCtraceFIM}.

%\textcolor{red}{We have to make sure that when we apply this MAC to BC transformation in sensing-related problems, it can be solved by assuming that the $i^{th}$ optimization variable is optimized while all are fixed. In other terms, we have to use block coordinate-related optimization techniques.}
    
\section{Problem Formulation and Solution for Multi-user Scenario}
%In this section, we formulate the optimization problem. 

Based on \eqref{MIMOBCCAP} and \eqref{traceJSimp}, we formulate a multi-objective optimization problem. Here, we focus on single-antenna multiple communication user scenarios with single sensing object. Moreover, in this part, we apply the BC to MAC transformation for further simplifications.

\subsection{Problem Formulation}
Our main objective is to maximize the weighted sum of the MI (sum-rate capacity) and the FI under the power budget constraint. The problem formulation for the MIMO BC is as follows:
%\vspace{1mm}
\begin{subequations}\label{BCmain1}
    \begin{align}
        \mathrm{P1}:\underset{\{{\mathbf{R}}_{i}\}_{i=1}^{K},{\mathbf{R}_0}} {\text{maximize}} & \quad \alpha \operatorname{tr}\left(\mathbf{M} \left(\sum_{i=1}^K\mathbf{R}_{i}+\mathbf{R}_0\right)\right) \nonumber \\& \quad + (1-\alpha) \sum_{i=1}^{K} \log{\left \vert 1+\mathbf{h}_{i}^H({\sum}_{j=1}^{i}\mathbf{R}_{j})\mathbf{h}_{i}\right \vert \over \left \vert 1+\mathbf{h}_{i}^H({\sum}_{j=1}^{i-1}\mathbf{R}_{j}) \mathbf{h}_{i}\right \vert }\\
        \text{subject to} && \nonumber \\
        &\mathrm{C}1:\sum_{i=1}^{K}\operatorname{tr}({\mathbf{R}}_{i}) + \operatorname{tr}(\mathbf{R}_0) \leq P_{tx} \label{mainBC_C1},\\
        &\mathrm{C}2: \mathbf{R}_{i} \succeq 0; \quad \forall i = 1,...,K, \label{mainBC_C2}\\
        & \mathrm{C}3: \mathbf{R}_{0} \succeq 0. \label{mainBC_C3}
        \end{align}
\end{subequations}

\noindent
\textit{{Proposition} 3.1:} With the condition of independently distributed of 
channels, the optimal solution of \eqref{BCmain1} satisfies: $\texttt{rank}(\mathbf{R}_i)=1, \; \forall i \in (1,..,K)$ and $\texttt{rank}(\mathbf{R}_0)=0$.

\noindent
\textit{Proof}: See \textbf{Appendix C}.  

Proposition 3.1 clearly shows that allocating a dedicated beam specifically for sensing is unnecessary to achieve the maximum weighted sum of the FI and the sum capacity as described in \eqref{BCmain1}. This finding implies that employing a joint beamforming strategy is preferable when dealing with multiple communication users with a sensing object rather than using separate beams for communication and sensing tasks.

With these findings, we can apply the uplink-downlink duality to the problem \eqref{BCmain1} with $\mathbf{R}_0 = 0$ for a more tractable problem formulation. Then, the new problem formulation is as follows:
\begin{subequations}\label{MainP1}
    \begin{align}
        \mathrm P2:\underset{\{{Q}_{i}\}_{i=1}^{K}}{\text{maximize}} &\quad \alpha \sum_{i=1}^K \operatorname{tr}(\mathbf{M}\mathbf{c}_{i}{Q}_{i}\mathbf{c}_{i}^H) \nonumber \\& \quad + (1-\alpha) \sum_{i=1}^{K}  \log_2{ \left \vert   \mathbf{I} + \mathbf{h}_{ei} {Q}_{i}\mathbf{h}_{ei}^H\right \vert}\\  
        \text{subject to}&&\nonumber\\
        &\mathrm{C}1:\sum_{i=1}^{K}\operatorname{tr}({Q}_{i}) \leq P_{tx} \label{mainP1_C1},\\
        &\mathrm{C}2: {Q}_{i} \geq 0; \quad \forall i = 1,...,K, \label{mainP1_C2}
    \end{align}
\end{subequations}
where $\alpha \geq 0$ is the weighted value. 

In our optimization problem, the objective function has two terms: the first term $\alpha \sum_{i=1}^K \operatorname{tr}(\mathbf{M}\mathbf{c}_{i} {Q}_{i}\mathbf{c}_{i}^H)$ is linear w.r.t. the variable ${Q}_{i}$. The second term $(1-\alpha) \sum_{i=1}^{K}  \log_2{ \left \vert   \mathbf{I} + \mathbf{h}_{ei} {Q}_{i}\mathbf{h}_{ei}^H \right \vert}$ includes a  logarithm of a determinant and $\mathbf{I} + \mathbf{h}_{ei}{Q}_{i}\mathbf{h}_{ei}^H$ is a linear function. Since a linear function is convex and the logarithm of a determinant is also convex\cite{boyd2004convex}, both terms are convex. The constraint \eqref{mainP1_C1} is convex due to linearity, while \eqref{mainP1_C2} implies the positive semi-definiteness of the variable and is also a convex constraint. Accordingly, both the objective function and the constraints are convex. Hence, \eqref{MainP1} is a convex optimization problem.

%\vspace{-90mm}
\subsection{Proposed Solution}
%\vspace{-30mm}
For some optimization problems, we can solve 
the problem using optimization techniques w.r.t. the corresponding \(i^{th}\) variable while all other variables are fixed and do so far all variables iteratively until convergence. This method is known as a block coordinate ascent or descent algorithm \cite{1412050}. Because \eqref{MainP1} is a convex maximization problem, we can use numerical tools like CVX to solve it with a block coordinate ascent algorithm \cite{cvx}. 

Given the convexity of the problem in \eqref{MainP1}, we can easily verify that the optimization problem satisfies Slater's condition, ensuring a zero duality gap (i.e., strong duality) \cite{boyd2004convex}. Consequently, we will analyze the Lagrangian expression, initially omitting \eqref{mainP1_C2}, as follows:
%\vspace{-10pt}
\begingroup
\setlength{\abovedisplayskip}{0pt}  % Space above displayed equations
\setlength{\belowdisplayskip}{15pt}  % Space below displayed equations
\begin{align}\label{LagP1_1}
    \mathcal{L}(\{{Q}_{i}\}_{i=1}^K) =&  \alpha \sum_{i=1}^K \operatorname{tr}(\mathbf{M}\mathbf{c}_{i}{Q}_{i}\mathbf{c}_{i}^H) - \beta \left(\sum_{i=1}^{K}\operatorname{tr}({Q}_{i}) - P_{tx}\right) \nonumber \\& \quad +
                (1-\alpha) \sum_{i=1}^{K}  \log_2{ \left \vert   \mathbf{I} + \mathbf{h}_{ei} {Q}_{i}\mathbf{h}_{ei}^H\right \vert}  ,
\end{align}
\endgroup
where $\beta \geq 0$ is the Lagrangian variable, and \eqref{LagP1_1} can be transformed as:
%\vspace{2pt}
\begingroup
\setlength{\abovedisplayskip}{10pt}  % Space above displayed equations
\setlength{\belowdisplayskip}{20pt}  % Space below displayed equations
\begin{align}\label{LagP1_2}
    \mathcal{L}(\{{Q}_{i}\}_{i=1}^K) =& \quad \beta P_{tx} - \sum_{i=1}^K \operatorname{tr}\left((\beta - \alpha \mathbf{c}_{i}^H\mathbf{M}\mathbf{c}_{i}){Q}_{i}\right) \nonumber \\& \quad + \left(\frac{1-\alpha}{\ln{2}}\right)\sum_{i=1}^{K} \ln{ \left \vert   \mathbf{I} + \mathbf{h}_{ei} {Q}_{i}\mathbf{h}_{ei}^H\right \vert}.
\end{align}
\endgroup
%\vspace{-10pt}

Let the optimal solution be denoted by ${Q}^{*}_{i}, \hspace{1mm}\forall {i}$. Moreover, to determine the optimal Lagrangian variable ($\beta^{*}$), any line search method can be effectively applied on a user-by-user basis. In \textbf{Appendix D}, we present the solution approach for the problem. We provide the detailed steps for implementing this solution approach in \textbf{Algorithm 1}.

\subsection{Analysis of Computational Complexity}
In \textbf{Algorithm 1}, we examine the optimization algorithm for the multi-user scenario. In Part 1, initializing \(\mathbf{c}_i\) has a complexity of \(\mathcal{O}(N_t^3)\) due to the required matrix inversion. Computing \(\mathbf{M}\) (as outlined in \eqref{traceJSimp}) also has a complexity of \(\mathcal{O}(N_t^3)\) because it involves matrix multiplication. The computational complexity of the third for loop (lines 10 to 17) is \(K \mathcal{O}(N_t^3)\), as it involves matrix inversion and depends on the number of communication users \(K\). Thus, the overall complexity of Part 1 is \((K+2) \mathcal{O}(N_t^3)\), which simplifies to \(K \mathcal{O}(N_t^3)\). Similarly, Part 2  has a simplified complexity of \(K \mathcal{O}(N_t^3)\). Therefore, the total computational complexity of \textbf{Algorithm 1} is \(2K \mathcal{O}(N_t^3)\).

%%%%%%%%%%%%%%%%%%%%%%%%%%%%%%%%%%%%%%%%%%%%%%%%%%%%
%\vspace{-10pt}
%\begin{figure*}[ht]
%\centering
%\begin{minipage}{\textwidth}

\begin{algorithm}[H]
\caption{Multi-objective Optimization of Transmit Covariance Matrix for Multi-user Scenario}
\label{algo:joint_op2}
%\vspace{10pt}
\text{Part 1 : Calculation of $\beta$ for given $P_{tx}$}\\
\textbf{Inputs:} $\mathbf{M}$, $P_{tx}$, $\beta_{final}$ \\
%\vspace{-10pt}
    \begin{algorithmic}[1]   
    \STATE $\beta_{\alpha} \gets []$ ; all $\beta$ values for each $\alpha$ 
        \FOR{$\alpha$ = 0 to 1} 
            \STATE $\mathbf{T}_f \gets [\ ]$ ; sum of covariance for each $\beta$
                \FOR{$\beta$= 1 to $\beta_{final}$}
                    \STATE Initialize $Q_{i} = 0$  for all $K$
                    \STATE Compute initial $\mathbf{T}_{i} = \mathbf{I}+\sum_{l=i+1}^K\left(\mathbf{h}_{l} {Q}_{l}\mathbf{h}_{l}^H\right)$
                    \STATE Compute initial ${S}_{i} = 1 + \mathbf{h}_{i}^H \left(\sum_{l=1}^{i-1}\mathbf{R}_{l}\right)\mathbf{h}_{i}$
                    \STATE Compute initial SVD of $\mathbf{T}_{i}^{-1/2}\mathbf{h}_{i}{S}_{i}^{1/2}$
                    \STATE Compute initial $\mathbf{c}_{i}^H = {S}_{i}^{1/2} \mathbf{f}_{i}^H\mathbf{T}_{i}^{-1/2}$ for all $K$
                    \FOR{$i$ = 1 to $K$}
                        \STATE Compute effective channel $\mathbf{h}_{ei}^H = \mathbf{h}_{i}^H\mathbf{A}_{M}^{-1/2}$ 
                        \STATE Compute \\ ${g}_{i} = \left(\mathbf{h}_{ei}^H \mathbf{h}_{ei}\right)^{1/2}\left(\beta - \alpha \mathbf{c}_{i}^H\mathbf{M}\mathbf{c}_{i}\right)^{-1/2} $
                        \STATE Compute ${g}_i = {u} \Sigma_{i} {v}$ \\ (Singular value decomposition)
                        \STATE Compute $\phi_i$ as in \ref{Lag_P1_sol}
                        \STATE Update $Q_f \gets Q_i$
                        \STATE Update (\ref{macbc_ci}) with $Q_i$
                    \ENDFOR
                    \RETURN $\mathbf{T}_f \gets \sum_{i=1}^K (Q_i)$
                \ENDFOR
                \STATE Plot $\beta$ vs $\mathbf{T}_f$ 
                \STATE Find the $\beta$ value corresponding to the $P_{tx}$
                \STATE Update $\beta_{\alpha} \gets \beta$
        \ENDFOR
        \RETURN $\beta_{\alpha}$ 
    \end{algorithmic}
    \vspace{10 pt}
\text{Part 2: Find the optimal uplink covariance}\\
\textbf{Inputs:} $\beta_{\alpha}$, $\mathbf{M}$, $P_{tx}$
\begin{algorithmic}[1]
    \FOR{$\alpha$ = 0 to 1}
        \STATE Initialize $Q_i$ = 0 for all $K$ , $\mathbf{Q}_f$
        \STATE Compute initial $\mathbf{T}_{i} = \mathbf{I}+\sum_{l=i+1}^K\left(\mathbf{h}_{l}{Q}_{l}\mathbf{h}_{l}^H\right)$
        \STATE Compute initial ${S}_{i} = 1 + \mathbf{h}_{i}^H \left(\sum_{l=1}^{i-1}\mathbf{R}_{l}\right)\mathbf{h}_{i}$
        \STATE Compute initial SVD of $\mathbf{T}_{i}^{-1/2}\mathbf{h}_{i} {S}_{i}^{1/2}$
        \STATE Compute initial $\mathbf{c}_{i}^H = {S}_{i}^{1/2}\mathbf{f}_{i}^H\mathbf{T}_{i}^{-1/2}$ for all $K$
        \FOR{$i$=1 to $K$}
            \STATE Compute effective channel $\mathbf{h}_{ei}^H = \mathbf{h}_{i}^H \mathbf{A}_{M}^{-1/2}$ 
            \STATE Compute \\ ${g}_{i}  = \left(\mathbf{h}_{ei}^H \mathbf{h}_{ei}\right)^{1/2}\left(\beta - \alpha \mathbf{c}_{i}^H\mathbf{M}\mathbf{c}_{i}\right)^{-1/2} $
            \STATE Compute ${g}_i = {u} \Sigma_{i} {v}$ \\ (Singular value decomposition)
            \STATE Compute $\phi_i$ as in (\ref{Lag_P1_sol})
        \ENDFOR
        \RETURN $Q_i$
    \ENDFOR
    \RETURN $\mathbf{Q}_f \gets Q_i \forall i$ 
    \STATE Compute $\mathbf{R}_x$ from  (\ref{macbc_ci})
\end{algorithmic}
\end{algorithm}

%\end{minipage}
%\end{figure*}
%%%%%%%%%%%%%%%%%%%%%%%%%%%%%%%%%%%%%%%%%%%%%%%%%%%%%%%%
%\vspace{-18pt}
\section{Problem Formulation and Solution for Single-user Scenario}
In this section, we examine the impact of EIRP on a JCAS-enabled MIMO system with a single downlink communication user and a single sensing object. EIRP is a crucial parameter for describing the transmitting antennas \cite{article}. It is essential in single-beam systems to ensure regulatory compliance, optimize coverage and signal quality, manage interference, enhance system efficiency, and improve sensing performance. However, the system treats EIRP as less critical in multi-user scenarios because it uses multiple beams, distributing the total transmit power among them.

\subsection{Problem formulation}
As  in Section III-A, we similarly define the objective function for this scenario as the weighted sum of the MI and the FI. We can state the optimization problem as follows:
\begin{subequations}
\label{Main_Sing}
  \begin{align}
    \underset{\mathbf R_{\mathbf{x}}}{\text{maximize}} &\quad \frac{2\alpha }{\sigma^2_r} \gamma_r^*\gamma_r \operatorname{tr}(\dot{\mathbf{A}} \mathbf{R}_{{x}} \dot{\mathbf{A}}^H ) +\;\frac{2\alpha }{\sigma^2_r}\operatorname{tr}(\mathbf{A} \mathbf{R}_{{x}} \mathbf{A}^H )  \nonumber
    \\& \quad +\;(1-\alpha)\log_2\left(1+ \frac{ \mathbf{h}^H \mathbf{R}_{x} \mathbf{h}}{ \sigma^2_{c}} \right) \\
    \text{subject to}&&\nonumber\\
    &\mathrm{C}1: \operatorname{tr}(\mathbf{R}_{{x}}) \leq P_{tx} \label{MainSing_C1}, \\
    &\mathrm{C}2: \mathbf{R}_{{x}} \preceq \text {EIRP} \label{MainSing_C2},\\
    &\mathrm{C}3: \mathbf{R}_{{x}} \succeq 0 \label{MainSing_C3}.
\end{align}  
\end{subequations}

Here, the optimization problem is convex due to linearity, and the constraints are convex due to linearity and the positive semi-definiteness of the variable matrix. 

\subsection{Proposed Solution}
As a result of the convexity of the optimization problem in \eqref{Main_Sing}, we can directly use the numerical tool, e.g., MATLAB CVX \cite{cvx}. However, instead of the numerical tools, we are using other optimization techniques to solve the problem while considering the CVX solution as a reference result to verify the accuracy of the techniques. Here, we mainly focus on the PGD method to solve the problem \cite{liu2007conjugate}, \cite{4641573}.
The gradient of the objective function in \eqref{Main_Sing} ($\mathbb{\nabla}(\text{f}(\mathbf{R}_\mathbf{x}))$) using Wirtinger derivatives is as follows:
\begin{equation}
    \label{DerivativePGD}
    \mathbb{\nabla}(\text{f}) = 2\alpha \mathbf{M}^{H}+2 \frac{(1-\alpha)}{\ln{2}}  \frac{\mathbf{K}^{H}}{(1+\operatorname{tr}(\mathbf{K}\mathbf{R}_{x}))},
\end{equation}
where $\mathbf{K} = \frac{\mathbf{h} \mathbf {h}^H}{\sigma^2_{c}}$. 
The gradient descent step for the covariance matrix $\mathbf{R}_\mathbf{x}$ is can be expressed as:
\begin{equation}
    \label{PD_NO_EIRP}
    \mathbf{R}^{\text {new}}_{x} = \mathbf{R}_{x} + {\delta} \mathbb{\nabla}(\text{f}),
\end{equation}
where the ${\delta}$ is the iteration-depend step-size.

As the next step of the solution approach, we use an orthogonal projection approach to do the projection onto the constraint set:
\begin{equation}
    \mathbf{C} = \mathbf{R}^{\text {new}}_{x} + \mathbf{E}_{o},
\end{equation}
where $\mathbf{E}_{o}$ is the distance between $\mathbf{C}$ and $\mathbf{R}^{\text {new}}_{x}$ and it can be found by
\begin{subequations}
\label{EProjection}
\begin{align}
    \mathrm P3:\underset{\mathbf E}{\text{minimize}} &\quad 
    \frac{1}{2}||\mathbf{C}-\mathbf{R}^{\text {new}}_{x}||^{2}_{\mathbf{F}} \\
   \text{subject to}&&\nonumber\\
    &\mathrm{C}1: \operatorname{tr}(\mathbf{C}) \leq P_{tx}, \label{pproj1} \\
    &\mathrm{C}2: (\text {EIRP}) \mathbf{I}- \mathbf{C} \succeq 0, \label{pproj2}\\
    &\mathrm{C}3: \mathbf{C} \succeq 0, \label{pproj3}
\end{align}
\end{subequations}
where $\mathbf{C}=\mathbf{U}\mathbf{\Lambda}\mathbf{U}^{H}$ and $\mathbf{R}^{\text{new}}_{x}=\mathbf{U}\mathbf{\Sigma}\mathbf{U}^{H}$ are the eigenvalue decomposition (EVD) for the problem due to both $\mathbf{C}$ and $\mathbf{R}^{\text {new}}_x$ is in the same eigen basis.
Therefore, the simplification of \eqref{EProjection} is 
\begin{subequations}
\label{ESProjection}
\begin{align}
    \mathrm P4:\underset{\mathbf E}{\text{minimize}} &\quad 
    \frac{1}{2}||\mathbf{\Lambda}-\mathbf{\Sigma}||^{2}_{\mathbf{F}} \\
   \text{subject to}&&\nonumber\\
    &\mathrm{C}1: \operatorname{tr}(\mathbf{\Lambda}) \leq P_{tx} \label{spproj1}, \\
    &\mathrm{C}2: \bf{0} \leq \mathbf{\Lambda} \leq (\text{EIRP})\mathbf{I}  \label{spproj2}.
\end{align}
\end{subequations}

To solve the problem in \eqref{ESProjection}, we can apply the KKT method. The corresponding Lagrange function is:
\begin{align}
\label{LAG_SING}  \mathcal{L}_e(\mathbf{X},\mathbf{Y},\mathbf{\Lambda},\mu)  =& \frac{1}{2}||\mathbf{\Lambda}-\mathbf{\Sigma}||^{2}_{\mathbf{F}} + \mu (\operatorname{tr}(\mathbf{\Lambda}) - P_{tx}) \nonumber \\& \quad  + \operatorname{tr}(\mathbf{X}^{H}(\mathbf{\Lambda}-(\text{EIRP})\mathbf{I} )) - \operatorname{tr}(\mathbf{Y}^{H}\mathbf{\Lambda}),
\end{align}
where $\mu$ is the Lagrangian multiplier and $\mathbf{X}$ and $\mathbf{Y}$ are the hermitian matrix associate with the constraint \eqref{spproj2} ($\mathbf{X} = \mathbf{X}^{H}$ and $\mathbf{Y} = \mathbf{Y}^{H}$). The derivative w.r.t. $\mathbf{\Lambda}$ is
\begin{align}
\label{KKT1}
    \frac{\partial \mathcal{L}_e}{\partial \mathbf{\Lambda}} = \mathbf{\Lambda} -\mathbf{\Sigma}+\mu \mathbf{I}+\mathbf{X}-\mathbf{Y} = 0 \\ \nonumber
    \mathbf{\Lambda}  = \mathbf{\Sigma} + \mathbf{Y} -\mathbf{X} -\mu \mathbf{I}.
\end{align}

The other KKT conditions are in \eqref{KKT11}-\eqref{KKT17} which are solved by using an iterative process:
\begin{align}
    \mu (\operatorname{tr}(\mathbf{\Lambda}) - P_{tx}) = 0 \label{KKT11}\\ 
    \operatorname{tr}(\mathbf{X}^{H}(\mathbf{\Lambda}-(\text{EIRP})\mathbf{I} )) = 0 \label{KKT12}\\
    \operatorname{tr}(\mathbf{Y}^{H}\mathbf{\Lambda}) = 0 \label{KKT13}\\
     \mu \geq 0 \label{KKT14}\\
    \operatorname{tr}(\mathbf{\Lambda}) - P_{tx} \leq 0\label{KKT15}\\
    \mathbf{X},\mathbf{Y} \succeq 0 \label{KKT16} \\
    0 \leq \mathbf{\Lambda} \leq (\text{EIRP})\mathbf{I}. \label{KKT17}
\end{align}

We present the overall algorithm in $\textbf{Algorithm 2}$. The algorithm runs until the stopping criteria as in \eqref{stop} is satisfied with $\epsilon \in \mathbb{R}_+$ and the maximum number of iterations $\text{Iter}_{\text max}$ is reached:
\begin{align}\label{stop}
    \mathbf{E}_{t} =& \quad \alpha \operatorname{tr}(\mathbf{M} \mathbf{R}_{x}^t)+\frac{(1-\alpha)}{\ln{2}} \ln{\left \vert 1+\operatorname{tr}(\mathbf{K}\mathbf{R}_{x}^t)\right \vert }.
\end{align}

%\begin{figure*}[ht]
%\centering
%\begin{minipage}{\textwidth}
\begin{algorithm}[H]
\caption{Multi-objective Optimization of Transmit Covariance Matrix for Single-user Scenario}
\label{algo:joint_op}
\mbox{\textbf{Inputs:} $\epsilon$, $P_{tx}$, $\delta$, $\text{Iter}_{\text{max}}$, $\mathbf{M}$, $\mathbf{K}$}
\vspace{-10pt}
\begin{algorithmic}[1]
\STATE Initialize $\mathbf{R}^{0}_\mathbf{x}$
\STATE $\mathbf{R}_{x}^\text{t}$ $=$ $\mathbf{R}^{0}_{x}$
\STATE Compute $\mathbf{E}_{t}$
\REPEAT
    \STATE Compute Gradient $\mathbb{\nabla}(\text{f})$ as in \ref{DerivativePGD}
    \STATE Compute $\mathbf{R}^{{t+1}}_{x}$ $=$ $\mathbf{R}_\mathbf{x}^{t} + \beta \mathbb{\nabla}(\text{f})$
    \STATE Compute $\mathbf{R}^{{t+1}}_{x} = \mathbf{U}\mathbf{\Sigma}\mathbf{U}^{H}$
    \STATE Set $\mathbf{\Sigma}$ into non-increasing order
    \STATE $\mathbf{\Lambda}$ $\leftarrow$ Output of the projection
    \STATE Compute $\mathbf{R}^{\text{proj}}_{x} = \mathbf{U}\mathbf{\Lambda}\mathbf{U}^{H}$
    \STATE Set $\mathbf{R}_{x}^{t+1}$ = $\mathbf{R}^{\text{proj}}_{x}$
    \STATE Set $\mathbf{E}_{t}$ $=$ $\mathbf{E}_{t+1}$
\UNTIL $|\mathbf{E}_{t+1} - \mathbf{E}_{t}|$ $<$ $\epsilon$
\RETURN $\mathbf{R}_{x}$
\end{algorithmic}
\end{algorithm}
%\end{minipage}
%\end{figure*}

\vspace{-20pt}
\subsection{Analysis of Computational Complexity}
Consider the optimization algorithm presented in \textbf{Algorithm 2}. The complexity for computing \(\mathbf{M}\) (as outlined in \eqref{traceJ}) and \(\mathbf{K}\) (as described in \eqref{DerivativePGD}) is \(\mathcal{O}(N_t^3)\) and \(\mathcal{O}(N_t^2)\), respectively, due to the operations involved, such as matrix-matrix multiplication and the vector outer product. Similarly, determining \(\mathbb{\nabla}(\text{f})\) and \(\mathbf{R}^{t+1}_{x}\) also incurs a complexity of \(\mathcal{O}(N_t^3)\) as these calculations involve matrix-matrix multiplications. The EVD of \(\mathbf{R}^{t+1}_{x}\) has a computational complexity of \(\mathcal{O}(N_t^3)\). In the projection step, solving the problem using the KKT conditions results in an overall complexity of \(\mathcal{O}(N_t^3)\). The calculation of \(\mathbf{R}^{\text{proj}}_{x}\) also requires \(\mathcal{O}(N_t^3)\) in terms of computational resources. As a result, the overall computational complexity of the algorithm is \(7 \mathcal{O}(N_t^3) + \mathcal{O}(N_t^2)\), which simplifies to \(7 \mathcal{O}(N_t^3)\). 

%Note that this represents the worst complexity case. 

\section{Simulation Results}
This section presents and analyzes the numerical results obtained for the case of multi-user and single-user communication.
%\vspace{-10pt}
\subsection{Simulation Parameters}
The parametric channel model \cite{10417011},\cite{8827589} is considered for the downlink users in the simulations. Under the system model, the channel response for the BS to $k^{th}$ downlink user can be defined as:
\begin{align}\label{BS_kuser}
    \mathbf{h}_{i} =& \quad \sqrt{P_{Loss}} \sum_{q=1}^{Q_p} c_{i,q} \mathbf{a}_{i}(\theta_{i,q});\quad i = 1,...,K,
\end{align}
where $Q_p$ is number of channel paths, $c_{i,q}$ denotes the channel path gain. Moreover, $\theta_{i,q}$ is an angle uniformly distributed over $[0,2\pi]$ while $P_{Loss}$ is the distance depends on path loss, and it is given by
\begin{align}
    P_{Loss} =& \quad \text{P}_0 \left(\frac{d}{d_0}\right)^{\eta},
\end{align}
where $d_0$ is the reference distance and $\eta$ is the path loss exponent. Table 1 shows the basic simulation parameters used for single and multi-user scenarios. For the simulations, we use MATLAB R2023a. For all simulations, we place the sensing object at a fixed location ($\theta_r = 0$).

\begin{table}[htbp]
\caption{Simulation parameters and values}
\begin{center}

% Increase the row spacing by a factor of 1.5 (adjust this as needed)
\renewcommand{\arraystretch}{1.5}

\begin{tabular}{|l|c|}
\hline
\textbf{Simulation Parameter} & {\textbf{Value}} \\
\hline
Number of Communication users & ${K}$ = 1,4,6,8\\
\hline
Transmit power & $P_{tx} $ = 30 dBm \\
\hline
EIRP & $\text{EIRP} $ = 27.78 - 29.54 dBm \\
\hline
Noise variance & $\sigma_c^2$ = $\sigma_r^2 $ = 0 dBm \\
\hline
Channel path numbers & $Q_p$ = 6 \\
\hline
Path-loss exponent & $\eta$ = 3.2 \\
\hline
Reference distance & $d_0$ = 1 m \\
\hline
Operating frequency & 3 GHz\\
\hline
Distance between two adjacent antennas & $\lambda/2$\\
\hline
Channel path gain & $c_{i,q} \sim \mathcal{C} \mathcal{N}(0,1)$\\
\hline
\end{tabular}

\renewcommand{\arraystretch}{1} % Reset to default after the table
\label{tab1}
\end{center}
\end{table}
\vspace{-10pt}
\subsection{Multi-User Communication Case}
This study focuses on the trade-off between communication and sensing for various scenarios. %We consider all types of interference and noises as AWGN for all simulations and derivations. 

%%_______________________________________________________________
%\vspace{-10pt}
\subsubsection{Joint Beamforming for Multiple Users and Single Sensing Object} \label{subsub1}
 
\begin{figure}[!t]
\centering\vspace{-5mm}
\includegraphics[scale = 0.5]{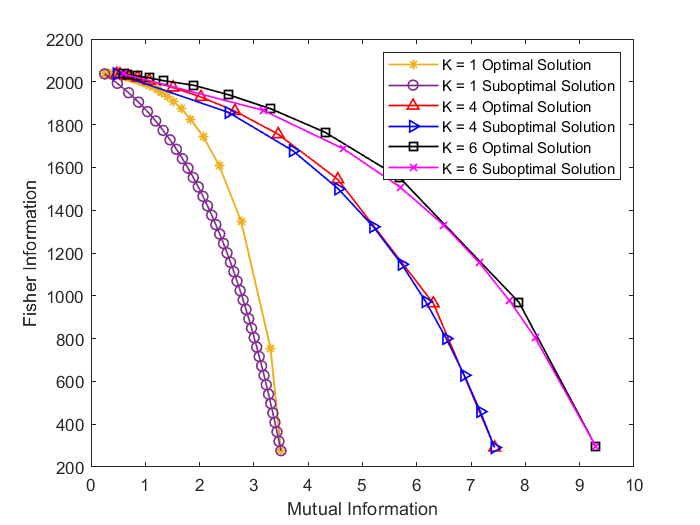}
\caption{Comparison between optimal and sub-optimal beamforming for various numbers of users in terms of the trade-off between sum rate and Fisher Information}
\label{FigMU1}
\vspace{-5mm}
\end{figure}

Fig. \ref{FigMU1} illustrates the performance of optimal and sub-optimal beamforming designs as the number of communication users varies, with \( N_t = N_r = 10 \). In optimal beamforming, there are \( K \) beams, whereas sub-optimal beamforming employs \( K+1 \) beams (as stated in \textbf{Proposition 3.1}). It is evident that as \( K \) increases, the gap between optimal and sub-optimal trade-offs decreases because increasing the number of communication users enhances system diversity and improves efficient resource allocation management, specifically transmit power. Consequently, the average system throughputs (MI and FI) for both beamforming designs converge to similar values, narrowing the performance gap between optimal and sub-optimal approaches. 

Moreover, as more users position themselves closer to the sensing object, the likelihood of covering the target increases, further reducing the gap between optimal and sub-optimal trade-offs. Therefore, with increasing numbers of communication users and constant \( N_t \) and \( N_r \), the optimal and sub-optimal scenarios approach each other more closely.

%==============================================================

\subsubsection{Joint Beamforming for Multiple Users and Single Sensing Object With Transmitting and Receiving Antenna Variation} \label{subsub2}
% Figure 
%%===================================================================
%%===================================================================

% Figure 
%%===================================================================
\begin{figure}[!t]
\centering\vspace{-5mm}
\includegraphics[scale = 0.5]{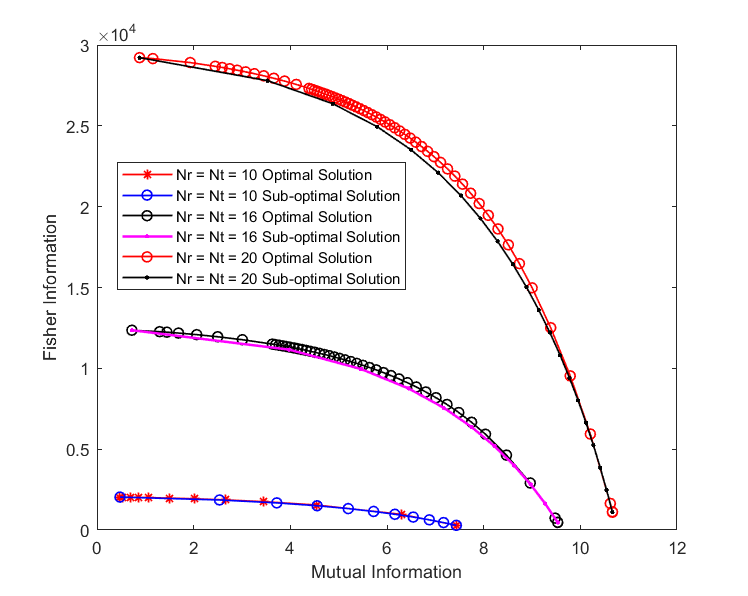}
\caption{Comparison between optimal and sub-optimal beamforming for various numbers of transmitting and receiving antennas in terms of the trade-off between sum-rate and Fisher Information}
\label{FigMU5}
\vspace{-5mm}
\end{figure}
%%===================================================================
We illustrate the optimal and sub-optimal beamforming design with variations in $N_r$ and $N_t$ for $K=4$. Fig.~\ref{FigMU5} demonstrates the optimal and sub-optimal variations for $N_r = N_t = 10$, $N_r = N_t = 16$, and $N_r = N_t = 20$. When we compare the gap between the optimal and sub-optimal results in these three scenarios, it is clear that the gap increases with increasing $N_r$ and $N_t$. As the number of transmit and receive antennas increases, the beamforming design becomes more complex with higher degrees of freedom (DoF). Moreover, interference management and resource allocation have become more complex with many antennas. Then, for larger systems, optimal performances become increasingly superior to the sub-optimal performances. Therefore, as shown in Fig.~\ref{FigMU5}, the gap between optimal and sub-optimal increases when $N_r$ and $N_t$ increase. 

From the observations in \ref{subsub1} and \ref{subsub2}, we can conclude that the case when the number of communication users increases with the constant number of transmitting and receiving antennas shows a weaker trade-off when compared to the trade-off with increasing number of transmitting and receiving antennas with the fixed number of communication users.
%==============================================================
\subsubsection{Variation in the Number of Communication Users}
% Figure 
%%===================================================================
\begin{figure}[!t]
\centering\vspace{0mm}
\includegraphics[scale = 0.5]{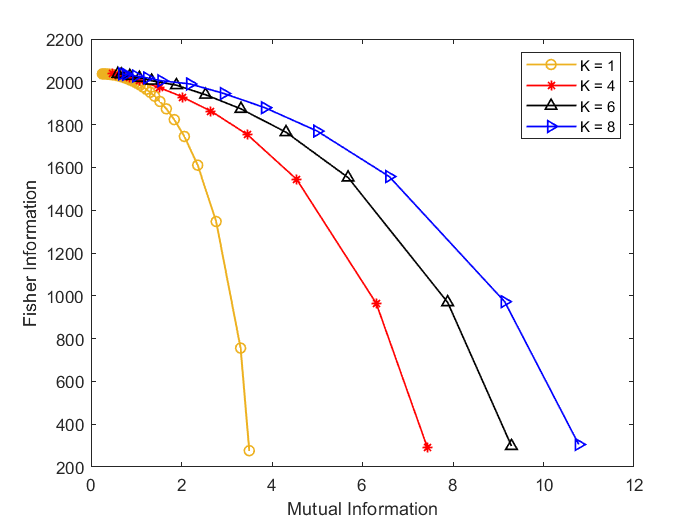}
\caption{Variation of trade-off with number of communication users}
\label{FigMU4}
\vspace{-5mm}
\end{figure}
%%===================================================================
We analyze the Variation of the trade-off with the increasing number of communication users. According to Fig. \ref{FigMU4} for \( N_r = N_t = 10 \), it is observed that the range of MI increases as \( K \) increases. With fixed total power allocation, increasing the number of communication users leads to a higher MI (total sum rate) due to enhanced spatial diversity, increased DoF, and improved interference management. 

Additionally, FI increases when the weighted value is 0. However, when the weighted value is 1, FI remains approximately constant due to increased communication users affecting the total FI, as shown in (\ref{eqn:FIM}). When \( \alpha = 1 \), the optimization prioritizes FI, and given only one sensing object, there can be no increase or decrease in sensing performance. Therefore, FI remains constant when \( \alpha = 1 \).

%==============================================================
\subsubsection{Transmit Power Budget Variation}
% Figure 
%%===================================================================
\begin{figure}[!t]
\centering\vspace{0mm}
\includegraphics[scale = 0.5]{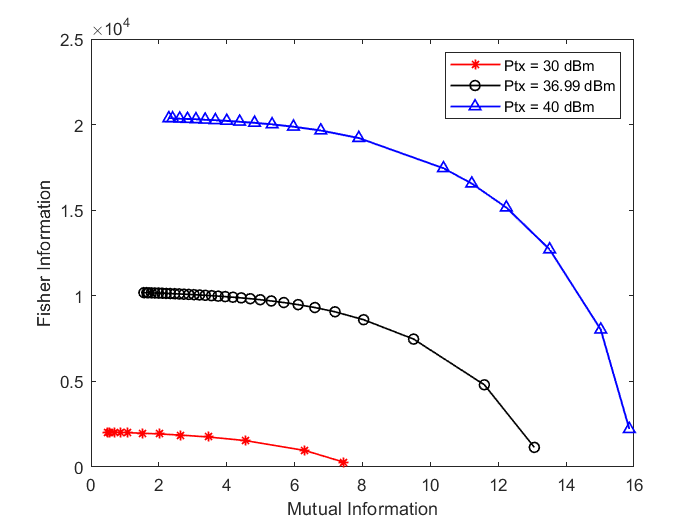}
\caption{Variation of trade-off with transmit power budget}
\label{FigMU3}
\vspace{-5mm}
\end{figure}
%%===================================================================

We explore how variations in transmit power affect the trade-off between communication and sensing performance. Increasing transmit power in our model enhances performance by improving the signal-to-noise ratio (SNR), facilitating higher data rates and more precise target parameter estimation. It also extends the effective range for communication and sensing, enabling coverage for more distant users and detecting objects at greater distances. As depicted in Fig. \ref{FigMU3}, higher \(P_{tx}\) levels result in enhanced performance for both communication and sensing while also widening the range of variability for both FI and MI.

%==============================================================
\subsubsection{Effect of Number of Transmitting Antennas}
% Figure 
%%===================================================================
\begin{figure}[!t]
\centering\vspace{0mm}
\includegraphics[scale = 0.5]{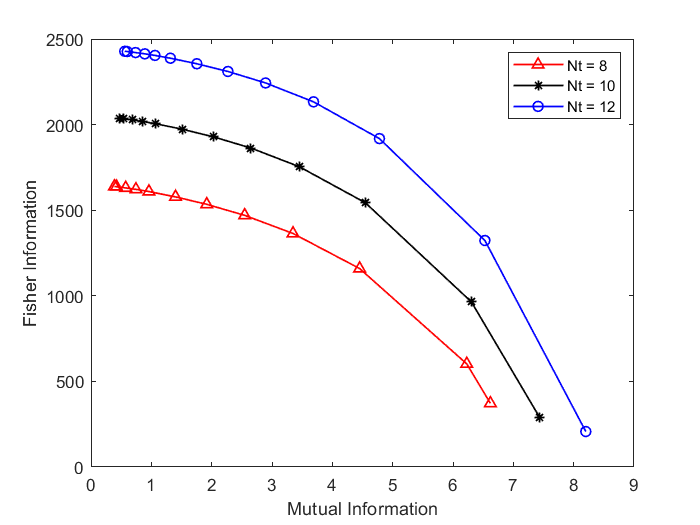}
\caption{Variation of trade-off with number of transmitting antennas}
\label{FigMU2}
\vspace{-5mm}
\end{figure}
%%===================================================================

We investigate how the number of transmitting antennas influences the trade-off between MI and FI. Increasing the number of transmitting antennas in a communication system enhances MI and FI. Additional antennas increase spatial diversity, reducing interference and improving signal quality. This enhancement results in higher data rates and better SNR through techniques like spatial multiplexing and beamforming, which increases MI. Simultaneously, more antennas improve spatial resolution and parameter estimation accuracy, such as direction of arrival and channel state information. The extra antennas also enable advanced processing techniques that reduce noise and interference, thereby improving FI.

Fig.~\ref{FigMU2} shows the trade-off between MI and FI for $N_r = 10$ and $K = 4$. When the weighted value is 0, we exclusively prioritize MI. At the same time, FI decreases because configurations that optimize MI may lead to higher interference and noise, which are detrimental to precise parameter estimation. When the weighted value is 1, MI approximately stabilizes, as the system's configurations that optimize FI do not necessarily support the highest possible data throughput while leading to enhanced sensing performance. This finding shows that increasing the number of transmitting antennas can simultaneously improve communication and sensing performances.

%==============================================================

\subsubsection{Variation in Number of Receiving Antennas}
% Figure 
%%===================================================================
\begin{figure}[!t]
\centering\vspace{0mm}
\includegraphics[scale = 0.5]{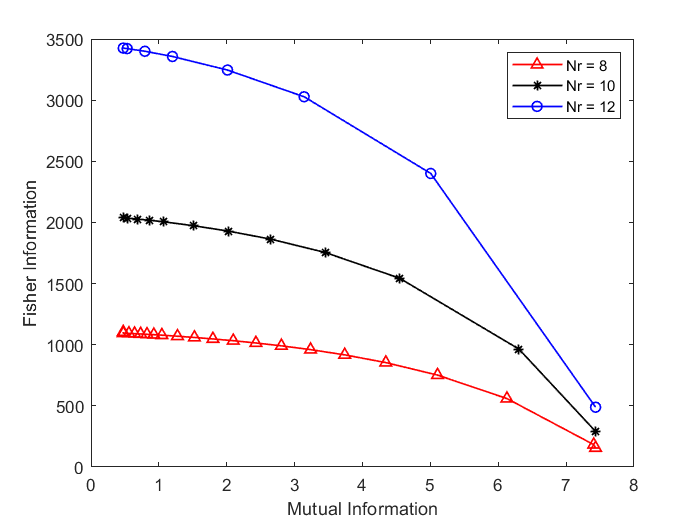}
\caption{Variation of trade-off with number of receiving antennas}
\label{FigMU33}
\vspace{-5mm}
\end{figure}
%%===================================================================
Fig.~\ref{FigMU33} demonstrates the trade-off with receiving antenna variation for 
$N_t = 10$ and $K=4$. Increasing the number of receiving antennas (\(N_r\)) in a MIMO system affects MI and FI differently. We can see that MI tends to reach a saturation range as $N_t$ increases, especially under fixed transmit power conditions. This saturation occurs due to the efficient utilization of spatial diversity and multiplexing gains alongside practical limitations such as noise and interference. Conversely, FI continuously improves with more receiving antennas, enhancing parameter estimation precision by providing additional independent observations of the received signal. This finding underscores that enhancing sensing performance can be achieved by increasing the number of receiving antennas.

%==============================================================
%==============================================================

\subsection{Single User Case With EIRP}
We consider $Nr = Nt = 4$ with a single communication user and sensing object. Furthermore, we change the EIRP value from 27.78 dBm to 29.54 dBm (i.e., from 0.6 W to 0.9 W). 

\subsubsection{Verification of Projected Gradient Descent}
% Figure 
%%===================================================================
\begin{figure}[!t]
\centering\vspace{0mm}
\includegraphics[scale = 0.5]{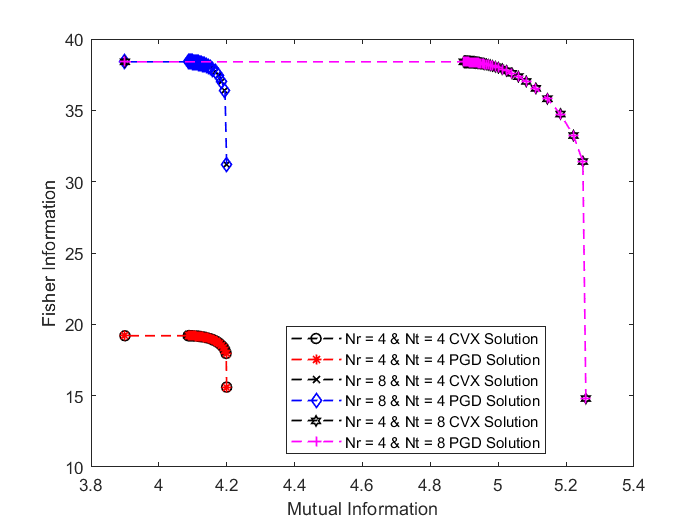}
\caption{Comparison between numerical and projected gradient descent (PGD) solutions for different $N_r$ and $N_t$ values.}
\label{FigSU1}
\vspace{-5mm}
\end{figure}
%%===================================================================
%The accuracy of the proposed algorithm is an important point. 
According to Fig.~\ref{FigSU1}, for EIRP is 27.78 dBm, shows the proposed optimization algorithm (see \textbf{Algorithm 1}) provides results similar to those from the numerical tool (i.e., MATLAB CVX). Moreover, we observe the behavior of the trade-off with the number of transmitting and receiving antenna variations in this EIRP scenario. When we increase the $N_r$, the sensing performances increase while the communication performances vary in a fixed range. On the other hand, when we increase $N_t$, it has a fixed sensing performance range while the communication performance increases. 

%==============================================================

\subsubsection{Variation of the Trade-off With EIRP Values}
% Figure 
%%===================================================================
\begin{figure}[!t]
\centering\vspace{0mm}
\includegraphics[scale = 0.5]{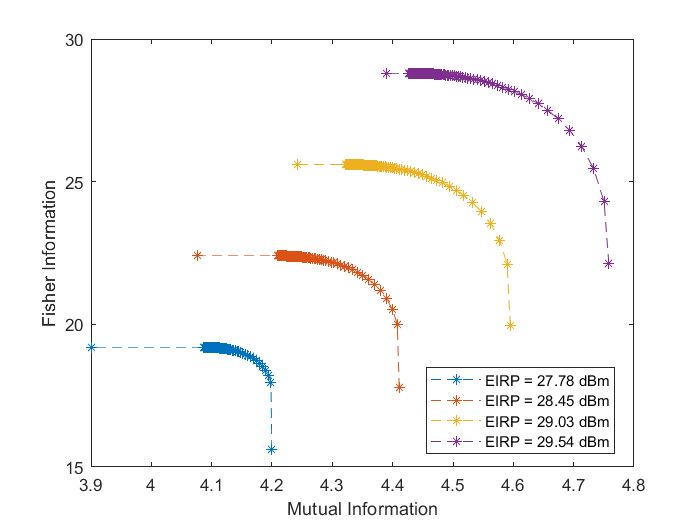}
\caption{Comparison between the communication and sensing performance trade-off with EIRP values.}
\label{FigSU2}
\vspace{-5mm}
\end{figure}
%%===================================================================
We illustrate the behavior of the trade-off between communication and sensing with varying EIRP values. Fig.~\ref{FigSU2} depicts this trade-off for various EIRP values. According to this figure, sensing and communication performances improve as the EIRP value increases. Furthermore, when the EIRP decreases, the points for $\alpha=0$ and $\alpha=1$ move away from the trade-off in the vertical and horizontal directions, respectively.

When $\alpha=0$, the optimization problem focuses on MI (Mutual Information), and theoretically, FI (Fisher Information) should be zero. Conversely, when $\alpha=1$, the focus is on FI, and MI should theoretically be zero. However, neither MI nor FI becomes zero when we optimize MI and FI in these two scenarios. Instead, they stabilize at specific values concerning the threshold value. Thus, for the cases of $\alpha=0$ and $\alpha=1$, FI and MI values can be placed between these vertical and horizontal straight lines. In conclusion, as EIRP decreases, the best trade-off is the curve within the $0 \leq \alpha \leq 1$.

\vspace{-3.5 pt}
\section{Conclusion}
%\vspace{-2 mm}
We have studied the trade-off between communication and sensing in MIMO JCAS over single-antenna multi-user, single-antenna single-user, and single-target scenarios. In particular, we have considered two scenarios: the multi-user case and the single-user case with EIRP constraints. For both cases, we have characterized the Pareto boundary by solving a multi-objective maximization optimization problem with FI and MI over the total transmit power budget and EIRP constraint via optimizing the transmit covariance matrix.
We have used the uplink-downlink duality conversion for the multi-user case to simplify the optimization problem and applied the block-coordinate ascent technique to solve it. The numerical results have revealed the communication and sensing trade-off. We have shown that joint beamforming is the optimal solution for JCAS, with the rank of the transmit covariance matrix equal to the number of communication users. Strong joint beamforming occurs with more transmitting and receiving antennas rather than a more significant number of communication users with a fixed number of antennas. We have illustrated the Pareto boundary using additional numerical examples and settings for this case. For the single-user case with EIRP limitations, we have shown that as the EIRP decreases, the optimal trade-off can be considered within the range of \(\alpha = (0,1)\). We can use the proposed model for JCAS to trade communication and sensing performances depending on the objective. We plan to examine a bi-static system model in future work, addressing interference separately and considering multiple sensing objects.

\section*{Appendix A}
\section*{Derivation of the Fisher Information Matrix}
This appendix describes the derivation method of the Fisher information used in \eqref{eqn:FIM}. With the assumption made in Section II-A, we consider the statistically independent complex Gaussian of the $\mathbf{y}_\mathbf{r}$. The PDF of the $\mathbf{y}_\mathbf{r}$ with the unknown deterministic parameters $\mathbf{v}=[\theta_r \quad \gamma_r \quad \gamma_r^{*}]$ can be defined as:

\begin{align}\label{pdf_1}
    p(\mathbf{y}_{r};\mathbf{v}) =& \quad \frac{1}{\left(\pi \sigma_r^2\right)^{N_t}} e^{-(\mathbf{y}_{r}-\gamma_r \mathbf{A}\mathbf{x}_{BC})^H({\sigma_r^2)}^{-1}(\mathbf{y}_{r}-\gamma_r \mathbf{A}\mathbf{x}_{BC})}.
\end{align}

As described in \cite{1703855}, \cite{4838872}, \cite{890346}, we can find the log-likelihood function for the PDF in \eqref{pdf_1} as mentioned in \eqref{logpdf} and the FI matrix ($\mathbf{J}$) as in \eqref{J_form}. Then, we can find the FI matrix concerning the ${i}^{th}$communication user as shown in \eqref{eqn:FIM}.
%\vspace{-5mm}
\begin{align}\label{logpdf}
    \ln{p(\mathbf{y}_{r};\mathbf{v})} =& \quad -N_t\ln{\left(\pi \sigma_r^2\right)} \\ \nonumber &\quad -\frac{(\mathbf{y}_{r}-\gamma_r \mathbf{A}\mathbf{x}_{BC})^H(\mathbf{y}_{r}-\gamma_r \mathbf{A}\mathbf{x}_{BC})}{\sigma_r^2}.
\end{align}

We can find the derivatives w.r.t. parameters in  $\mathbf{v}$ as follows:
\begin{align} \label{PD1}
    \frac{\partial \ln(p)}{\partial\theta_r} =& \quad \frac{-1}{\sigma_r^2} \{-\mathbf{y}_{r} \dot{\mathbf{A}}\mathbf{x}_{BC}\gamma_r + \gamma_r^H \gamma_r \mathbf{x}_{BC}^H \mathbf{A}\dot{\mathbf{A}} \mathbf{x}_{BC} \\ \nonumber &\quad  -\gamma_r^H \mathbf{x}_{BC}^H\dot{\mathbf{A}}^H\mathbf{y}_{r} + \gamma_r^H \mathbf{x}_{BC}^H\dot{\mathbf{A}}^H \mathbf{A}\mathbf{x}_{BC}\gamma_r\},
\end{align}

\begin{align} \label{PD2}
    \frac{\partial \ln(p)}{\partial\gamma_r} =& \quad \frac{1}{\sigma_r^2} \{(\mathbf{y}_{r}-\mathbf{A}\mathbf{x}_{BC}\gamma_r)^H\mathbf{A}\mathbf{x}_{BC}\},
\end{align}

\begin{align} \label{PD3}
    \frac{\partial \ln(p)}{\partial\gamma_r^*} =& \quad \frac{1}{\sigma_r^2} \{(\mathbf{A}\mathbf{x}_{BC})^H(\mathbf{y}_{r}-\mathbf{A}\mathbf{x}_{BC}\gamma_r)\}.
\end{align}

From (\ref{PD1}),  (\ref{PD2}), and (\ref{PD3}), we can derive the Fisher information matrix using equation (\ref{J_form}) below:

\begin{align}\label{J_form}
    \mathbf{J} = & \quad \mathbb{E}\left(\left(\frac{\partial(\ln{p(\mathbf{y}_{r};\mathbf{v})})}{\partial\mathbf{v}}\right) \left(\frac{\partial(\ln{p(\mathbf{y}_{r};\mathbf{v})})}{\partial\mathbf{v}}\right)^T\right).
\end{align}

\section*{Appendix B}
\section*{MAC to BC Transformation}
Here, we explain the relationship between the downlink beamforming matrix and the uplink beamforming matrix. This transformation is performed on a user-by-user basis, ensuring equal uplink and downlink sum-rate capacity as indicated in \eqref{BCMACEQUAL} \cite{1237143}, \cite{1412050}. Because the power budget is identical for both uplink and downlink:

\begin{equation}
    \sum_{i=1}^K {Q}_{i} = \sum_{i=1}^K \operatorname{tr}(\mathbf{R}_{i}).
\end{equation}

The equivalent downlink covariance matrix $\mathbf{R}_\mathbf{i}$ can be expressed as:
\begin{equation}\label{MACBCtransf}
    \mathbf{R}_{i} = \mathbf{T}_{i}^{-1/2}\mathbf{f}_{i}{J}_{i}{S}_{i}^{1/2}{Q}_{i}{S}_{i}^{1/2}{J}_{i}\mathbf{f}_{i}^{H}\mathbf{T}_{i}^{-1/2},
\end{equation}
where $\mathbf{T}_{i} = \mathbf{I}+\sum_{l=i+1}^K\left(\mathbf{h}_{l}{Q}_{l}\mathbf{h}_{l}^H\right)$ and ${S}_{i} = 1 + \mathbf{h}_{i}^H \left(\sum_{l=1}^{i-1}\mathbf{R}_{l}\right)\mathbf{h}_{i}$. Furthermore, $\mathbf{T}_{i}^{-1/2}\mathbf{h}_{i} {S}_{i}^{1/2}$ is a vector and SVD of this is straightforward, since ${J}_{i} = 1$. Then, with the simplified $\text{svd}\left(\mathbf{T}_{i}^{-1/2}\mathbf{h}_{i}{S}_{i}^{1/2}\right) = \mathbf{f}_{i} \text{D}_{i}$. For the transformation, we have to start with $i = 1$.
For simplicity of the equation we use the $\mathbf{c}_{i} \in \mathbb{C}^{1 \times N_t}$ as follows:
\begin{equation} \label{macbc_ci}
    \mathbf{c}_{i} = {S}_{i}^{1/2}\mathbf{f}_{i}^{H}\mathbf{T}_{i}^{-1/2}.
\end{equation}
%\vspace{-40pt}
\section*{Appendix C}
\section*{ Proof of Proposition 3.1}
The main optimization problem for BC in \eqref{BCmain1} is a convex optimization problem. Therefore, Slater's conditions verify a zero duality gap \cite{boyd2004convex}, \cite{6860253}. Thus, we consider the Lagrangian formulation for \eqref{BCmain1} as follows:
%\begin{equations}\label{BCLag1}
    \begin{align}\label{BCLag1}
        \mathcal{L}\left(\{{\mathbf{R}}_{i}\}_{i=1}^K , \mathbf{R}_0, \beta\right) =& \quad \delta P_{tx} + (1-\alpha){C_{BC}}^{'} + \sum_{i=1}^K \operatorname{tr}(\mathbf{R}_i^{'})\nonumber\\&
        +\operatorname{tr}(\mathbf{R}_0^{'}),
    \end{align}
%\end{equations}
where $\mathbf{R}_i^{'}=(\alpha\mathbf{M}-\delta\mathbf{I})^{1/2}\mathbf{R}_i(\alpha\mathbf{M}-\delta\mathbf{I})^{1/2}$, $\mathbf{R}_0^{'}=(\alpha\mathbf{M}-\delta\mathbf{I})^{1/2}\mathbf{R}_0(\alpha\mathbf{M}-\delta\mathbf{I})^{1/2}$ and $C_{BC}^{'} = \sum_{i=1}^{K} \log{\left \vert 1+\mathbf{y}_{i}^H({\sum}_{j=1}^{i}\mathbf{R}_{j}^{'})\mathbf{y}_{i}\right \vert \over \left \vert 1+\mathbf{y}_{i}^H({\sum}_{j=1}^{i-1}\mathbf{R}_{j}^{'}) \mathbf{y}_{i}\right \vert }$ with $\mathbf{y}_i = \mathbf{h}_i^H(\alpha\mathbf{M}-\delta\mathbf{I})^{-1/2}$. $\delta$ is the Lagrangian multiplier with $\delta \geq 0$.
The dual problem of the \eqref{BCLag1} is given as follows:
%\begin{equations}\label{dualLag1}
\begin{align}\label{dualLag1}
   \mathrm{\eqref{BCLag1}D}:\underset{\delta \geq 0} {\text{minimize}} & \quad \delta P_{tx} + (1-\alpha)C_{BC}^{'}.
\end{align}
%\end{equations}

The optimal solution for \eqref{dualLag1} is $\delta^{}$, and the resulting solutions are $\mathbf{R}_i^{'}$ and $\mathbf{R}_0^{'*}$. Then, the optimal solutions for \eqref{BCLag1} and \eqref{BCLag1}D should follow the complementary slackness conditions:
\begin{align}\label{slack1}
        \operatorname{tr}(\mathbf{R}_i^{'*}) = 0, \hspace{1mm} \forall i \in (1,...,K),       
\end{align}
%\vspace{-10 mm}
\begin{align}\label{slack2}
        \operatorname{tr}(\mathbf{R}_0^{'*})=0,       
\end{align}
which are equivalent to $\mathbf{R}_i^{'*}=0, \hspace{1mm}\forall i \in (1,...,K)$ and $\mathbf{R}_0^{'*} = 0$, respectively. To satisfy the SINR requirements, it is confirmed that $\mathbf{R}_i^{*} \neq 0$. Therefore, from the $\mathbf{R}_i^{'*}$ we can verify that $\texttt{rank}(\mathbf{R}_i) \geq 1, \hspace{1 mm} \forall i \in i,...,K$ and from \eqref{slack1}, $\texttt{rank}(\alpha\mathbf{M}-\delta^{*} \mathbf{I})\leq N_t -1$. Next, we will prove the $\mathbf{R}_0^{*}=0$. With the $(\alpha\mathbf{M}-\delta^{*} \mathbf{I}) \succeq 0$, $\mathbf{R}_0$ must lie in the null space of $(\alpha\mathbf{M}-\delta^{*} \mathbf{I})$. Therefore, $\mathbf{R}_0^{*} = 0$ with $\texttt{rank}(\alpha\mathbf{M}-\delta^{*} \mathbf{I}) \geq N_t -1$. From that point we can sat for any $\mathbf{R}_i^{*} \succeq 0$ satisfying \eqref{slack2} and it must hold that $\mathbf{R}_0^{*} = 0$. Finally, with the $\texttt{rank}(\alpha\mathbf{M}-\delta^{*} \mathbf{I})\leq N_t -1$ and $\texttt{rank}(\alpha\mathbf{M}-\delta^{*} \mathbf{I}) \geq N_t -1$, it follows that the $\texttt{rank}(\alpha\mathbf{M}-\delta^{*} \mathbf{I}) = N_t -1$. Therefore, with the rank properties, we can be concluded that the $\texttt{rank}(\mathbf{R}_i) = 1, \hspace{1 mm} \forall I \in I,..., K$. Therefore, Proposition 3.1 is proven.

\section*{Appendix D}
\section*{Solution for the Multi-user Case}
By using the properties of determinant ($\text{det}(\mathbf{I}+\mathbf{u}\mathbf{v}^{H})=1+\mathbf{u}^H\mathbf{v}$)~\cite{Petersen2008}, we can reformulate the Lagrangian formulation mentioned in \eqref{LagP1_2} as:

\begin{align} \label{Lag_P1_3}
    \mathcal{L}\left(\{{Q}^{'}_{i}\}_{i=1}^K, \beta\right) =& \quad \beta P_{tx} -  \sum_{i=1}^K \operatorname{tr}({Q}^{'}_{i}) \nonumber \\& \quad + \left(\frac{1-\alpha}{\ln{2}}\right)\sum_{i=1}^{K} \ln{ \left (  1 + {g}_{i} {Q}^{'}_{i} {g}_{i}\right )},
\end{align}
where ${Q}^{'}_{i} = \left(\beta - \alpha \mathbf{c}_{i}^H\mathbf{M}\mathbf{c}_{i}\right)^{1/2}{Q}_{i}\left(\beta - \alpha \mathbf{c}_{i}^H\mathbf{M}\mathbf{c}_{i}\right)^{1/2} \in \mathbb{C} $ and  ${g}_{i} = \left(\mathbf{h}_{ei}^H \mathbf{h}_{ei}\right)^{1/2}\left(\beta - \alpha \mathbf{c}_{i}^H \mathbf{M}\mathbf{c}_{i}\right)^{-1/2} \in \mathbb{C}$.

By applying the SVD for ${g}_{i}$, $\text{svd}({g}_{i}) = {u} \Sigma_{i} {v}$\footnote{Note that when using MATLAB to solve the problem, have to careful to use the SVD command with 'econ' to produces a different economy-size decomposition.}, and  ${Q}^{'}_{i} = {v} \phi_{i} {v}$, the Lagrangian function is:

\begin{align}\label{Lag_P1_4}
    \mathcal{L}\left(\{{Q}^{'}_{i}\}_{i=1}^K\right) =& \quad \beta P_{tx} - \sum_{i=1}^K \operatorname{tr}(\phi_{i}) \nonumber \\& \quad + \left(\frac{1-\alpha}{\ln{2}}\right)\sum_{i=1}^{K} \ln{\left(1+\Sigma_{i}\phi_{i}\Sigma_{i}\right)}.
\end{align}

The derivative w.r.t. the complex variable (Wirtinger derivatives) $\phi_{i}$ is \cite{Petersen2008},\cite{Haykin:2002}:

\begin{align}\label{Lag_P1_deri}
    \frac{\partial \mathcal{L}}{\partial\phi_{i}} =& 2\left(\frac{1-\alpha}{\ln{2}}\right)\Sigma_{i}\left(1+\Sigma_{i}\phi_{i}\Sigma_{i}\right)^{-1}\Sigma_{i}- (2 \times 1).
\end{align}

By solving the \eqref{Lag_P1_deri}, we can find a optimal solution for $\phi^{*}_{i}$ as follows:
\begin{align}\label{Lag_P1_sol}
    \phi^{*}_{i} =& \quad \left(\frac{1-\alpha}{\ln{2}}\right) - \frac{1}{\Sigma_{i}^{2}}.
\end{align}
%Here, we can see the value for $\alpha$ and $\beta$ $\phi_{i} > 0$. Therefore, rank($\phi^{*}_{i}$) = 1.

Let us consider the MAC to BC transformation to find the optimal downlink beamforming matrix ($\mathbf{R}^{*}_{i}$) w.r.t. $\alpha$ and $\beta$ from \eqref{MACBCtransf} and \eqref{Lag_P1_sol}:

\begin{align}\label{MACBCtransf_2}
    \mathbf{R}^{*}_{i} =& \quad \mathbf{T}_{i}^{-1/2}\mathbf{f}_{i}{J}_{i} {S}_{i}^{1/2} \left(\beta - \alpha \mathbf{c}_{i}^H\mathbf{M}\mathbf{c}_{i}\right)^{-1/2} {v} \phi^{*}_{i} {v} \\ \nonumber &\quad \left(\beta - \alpha \mathbf{c}_{i}^H\mathbf{M}\mathbf{c}_{i}\right)^{-1/2}{S}_{i}^{1/2}{J}_{i}\mathbf{f}_{i}^{H}\mathbf{T}_{i}^{-1/2}.
\end{align}

%Since $\text{rank}(\mathbf{AB}) \leq \text{min}(\text{rank}(\mathbf{A}),\text{rank}(\mathbf{B}))$, we can prove that  $\text{rank}(\mathbf{R}^{*}_{i}) \leq \text{min}(N_t,1)$. Therefore, it can be proved that  $\text{rank}(\mathbf{R}^{*}_{i}) = 1, \; \forall i$, while $\mathbf{R}^{*}_\mathbf{x} = \sum_{i=1}^{K}\mathbf{R}^{*}_{i}$. Due to linear independence of each $\mathbf{R}^{*}_{i}, \; \forall i$, using the rank properties, $\text{rank}(\mathbf{R}^{*}_\mathbf{x}) = K $. Thus, \textbf{Proposition 3.1} is proved.

% -------------------------------------------------------------------------
%\vspace{-5mm}
\bibliographystyle{IEEEtran}
\bibliography{ref}

@ARTICLE{1412050,
  author={Jindal, N. and Wonjong Rhee and Vishwanath, S. and Jafar, S.A. and Goldsmith, A.},
  journal={IEEE Transactions on Information Theory}, 
  title={Sum power iterative water-filling for multi-antenna Gaussian broadcast channels}, 
  year={2005},
  volume={51},
  number={4},
  pages={1570-1580},
  doi={10.1109/TIT.2005.844082}}

@misc{liu2007conjugate,
      title={Conjugate Gradient Projection Approach for Multi-Antenna Gaussian Broadcast Channels}, 
      author={Jia Liu and Y. Thomas Hou and Hanif D. Sherali},
      year={2007},
      eprint={cs/0701061},
      archivePrefix={arXiv},
      primaryClass={cs.IT}
}

@misc{li2023framework,
      title={{A Framework for Mutual Information-based MIMO Integrated Sensing and Communication Beamforming Design}}, 
      author={Jin Li and Gui Zhou and Tantao Gong and Nan Liu},
      year={2023},
      eprint={2211.07887},
      archivePrefix={arXiv},
      primaryClass={eess.SP}
}

@ARTICLE{1327794,
  author={Wei Yu and Cioffi, J.M.},
  journal={IEEE Transactions on Information Theory}, 
  title={Sum capacity of Gaussian vector broadcast channels}, 
  year={2004},
  volume={50},
  number={9},
  pages={1875-1892},
  doi={10.1109/TIT.2004.833336}}

@ARTICLE{1207369,
  author={Caire, G. and Shamai, S.},
  journal={IEEE Transactions on Information Theory}, 
  title={On the achievable throughput of a multiantenna Gaussian broadcast channel}, 
  year={2003},
  volume={49},
  number={7},
  pages={1691-1706},
  doi={10.1109/TIT.2003.813523}}

@ARTICLE{1056659,
  author={Costa, M.},
  journal={IEEE Transactions on Information Theory}, 
  title={Writing on dirty paper (Corresp.)}, 
  year={1983},
  volume={29},
  number={3},
  pages={439-441},
  doi={10.1109/TIT.1983.1056659}}

@ARTICLE{1237143,
  author={Vishwanath, S. and Jindal, N. and Goldsmith, A.},
  journal={IEEE Transactions on Information Theory}, 
  title={Duality, achievable rates, and sum-rate capacity of Gaussian MIMO broadcast channels}, 
  year={2003},
  volume={49},
  number={10},
  pages={2658-2668},
  doi={10.1109/TIT.2003.817421}}

@book{kay1993fundamentals,
  title={{Fundamentals of statistical signal processing: estimation theory}},
  author={Kay, Steven M},
  year={1993},
  publisher={Prentice-Hall, Inc.}
}

@misc{cvx,
  author       = {CVX Research, Inc.},
  title        = {{CVX}: Matlab Software for Disciplined Convex Programming, version 2.0},
  howpublished = {\url{https://cvxr.com/cvx}},
  month        = aug,
  year         = 2012
}

@book{boyd2004convex,
  title={Convex Optimization},
  author={Boyd, S.P. and Vandenberghe, L.},
  year={2004},
  publisher={Cambridge University Press}
}

@misc{Petersen2008,
  author = {Petersen, K. B. and Pedersen, M. S.},
  title = {The Matrix Cookbook},
  year = 2008
}

@book{Haykin:2002,
  address = {Upper Saddle River, NJ},
  author = {Haykin, Simon},
  description = {diverse cognitive systems bib},
  edition = {4th},
  publisher = {Prentice Hall},
  title = {Adaptive filter theory},
  year = 2002
}

@ARTICLE{1703855,
  author={Bekkerman, I. and Tabrikian, J.},
  journal={IEEE Transactions on Signal Processing}, 
  title={Target Detection and Localization Using MIMO Radars and Sonars}, 
  year={2006},
  volume={54},
  number={10},
  pages={3873-3883}}

@ARTICLE{4838872,
  author={Ben-Haim, Zvika and Eldar, Yonina C.},
  journal={IEEE Signal Processing Letters}, 
  title={On the Constrained CramÉr–Rao Bound With a Singular Fisher Information Matrix}, 
  year={2009},
  volume={16},
  number={6},
  pages={453-456}}

@ARTICLE{890346,
  author={Stoica, P. and Marzetta, T.L.},
  journal={IEEE Transactions on Signal Processing}, 
  title={Parameter estimation problems with singular information matrices}, 
  year={2001},
  volume={49},
  number={1},
  pages={87-90}}

@INPROCEEDINGS{4641573,
  author={Hunger, Raphael and Schmidt, David A. and Joham, Michael and Utschick, Wolfgang},
  booktitle={2008 IEEE 9th Workshop on Signal Processing Advances in Wireless Communications}, 
  title={A general covariance-based optimization framework using orthogonal projections}, 
  year={2008},
  volume={},
  number={},
  pages={76-80}}

@ARTICLE{10417011,
  author={Wijekoon, Dilki and Mezghani, Amine and Hossain, Ekram},
  journal={IEEE Transactions on Wireless Communications}, 
  title={Phase Shifter Optimization in RIS-Aided MIMO Systems Under Multiple Reflections}, 
  year={2024},
  volume={},
  number={},
  pages={1-1}}

@ARTICLE{8827589,
  author={Rahman, Md. Lushanur and Zhang, J. Andrew and Huang, Xiaojing and Guo, Y. Jay and Heath, Robert W.},
  journal={IEEE Transactions on Aerospace and Electronic Systems}, 
  title={Framework for a Perceptive Mobile Network Using Joint Communication and Radar Sensing}, 
  year={2020},
  volume={56},
  number={3},
  pages={1926-1941}}

@article{b2d30f91f67f4698946766bd16786c74,
title = "5G, 6G, and Beyond: Recent advances and future challenges",
author = "Fatima Salahdine and Tao Han and Ning Zhang",
year = "2023",
month = oct,
doi = "10.1007/s12243-022-00938-3",
volume = "78",
pages = "525--549",
journal = "Annales des Telecommunications/Annals of Telecommunications",
issn = "0003-4347",
publisher = "Springer Paris",
number = "9-10",
}

@ARTICLE{9354629,
  author={Wild, Thorsten and Braun, Volker and Viswanathan, Harish},
  journal={IEEE Access}, 
  title={Joint Design of Communication and Sensing for Beyond 5G and 6G Systems}, 
  year={2021},
  volume={9},
  number={},
  pages={30845-30857},
  doi={10.1109/ACCESS.2021.3059488}}

@INPROCEEDINGS{8905229,
  author={Rahman, Md Lushanur and Cui, Peng-fei and Zhang, J. Andrew and Huang, Xiaojing and Guo, Y. Jay and Lu, Zhiping},
  booktitle={2019 19th International Symposium on Communications and Information Technologies (ISCIT)}, 
  title={Joint Communication and Radar Sensing in 5G Mobile Network by Compressive Sensing}, 
  year={2019},
  volume={},
  number={},
  pages={599-604},
  doi={10.1109/ISCIT.2019.8905229}}

@INPROCEEDINGS{9843013,
  author={Decarli, Nicolò and Bartoletti, Stefania and Masini, Barbara M.},
  booktitle={2022 IEEE 21st Mediterranean Electrotechnical Conference (MELECON)}, 
  title={Joint Communication and Sensing in 5G-V2X Vehicular Networks}, 
  year={2022},
  volume={},
  number={},
  pages={295-300},
  doi={10.1109/MELECON53508.2022.9843013}}

@misc{pucci2022systemlevel,
      title={System-Level Analysis of Joint Sensing and Communication based on 5G New Radio}, 
      author={Lorenzo Pucci and Enrico Paolini and Andrea Giorgetti},
      year={2022},
      eprint={2201.12017},
      archivePrefix={arXiv},
      primaryClass={eess.SP}
}

@INPROCEEDINGS{10008554,
  author={Liu, Meng and Yang, Minglei and Nallanathan, Arumugam},
  booktitle={2022 IEEE Globecom Workshops (GC Wkshps)}, 
  title={On the Performance of Uplink and Downlink Integrated Sensing and Communication Systems}, 
  year={2022},
  volume={},
  number={},
  pages={1236-1241},
  doi={10.1109/GCWkshps56602.2022.10008554}}

@ARTICLE{9737357,
  author={Liu, Fan and Cui, Yuanhao and Masouros, Christos and Xu, Jie and Han, Tony Xiao and Eldar, Yonina C. and Buzzi, Stefano},
  journal={IEEE Journal on Selected Areas in Communications}, 
  title={Integrated Sensing and Communications: Toward Dual-Functional Wireless Networks for 6G and Beyond}, 
  year={2022},
  volume={40},
  number={6},
  pages={1728-1767},
  doi={10.1109/JSAC.2022.3156632}}

@INPROCEEDINGS{9764187,
  author={Liu, Fan and Liu, Ya-Feng and Masouros, Christos and Li, Ang and Eldar, Yonina C.},
  booktitle={2022 IEEE Radar Conference (RadarConf22)}, 
  title={A Joint Radar-Communication Precoding Design Based on Cramér-Rao Bound Optimization}, 
  year={2022},
  volume={},
  number={},
  pages={1-6},
  doi={10.1109/RadarConf2248738.2022.9764187}}

@article{Liu_2022,
   title={Cramér-Rao Bound Optimization for Joint Radar-Communication Beamforming},
   volume={70},
   ISSN={1941-0476},
   url={http://dx.doi.org/10.1109/TSP.2021.3135692},
   DOI={10.1109/tsp.2021.3135692},
   journal={IEEE Transactions on Signal Processing},
   publisher={Institute of Electrical and Electronics Engineers (IEEE)},
   author={Liu, Fan and Liu, Ya-Feng and Li, Ang and Masouros, Christos and Eldar, Yonina C.},
   year={2022},
   pages={240–253} }

@misc{liu2021surveyfundamentallimitsintegrated,
      title={A Survey on Fundamental Limits of Integrated Sensing and Communication}, 
      author={An Liu and Zhe Huang and Min Li and Yubo Wan and Wenrui Li and Tony Xiao Han and Chenchen Liu and Rui Du and Danny Tan Kai Pin and Jianmin Lu and Yuan Shen and Fabiola Colone and Kevin Chetty},
      year={2021},
      eprint={2104.09954},
      archivePrefix={arXiv},
      primaryClass={cs.IT},
      url={https://arxiv.org/abs/2104.09954}, 
}

@ARTICLE{10251151,
  author={Ren, Zixiang and Peng, Yunfei and Song, Xianxin and Fang, Yuan and Qiu, Ling and Liu, Liang and Ng, Derrick Wing Kwan and Xu, Jie},
  journal={IEEE Transactions on Wireless Communications}, 
  title={Fundamental CRB-Rate Tradeoff in Multi-Antenna ISAC Systems With Information Multicasting and Multi-Target Sensing}, 
  year={2024},
  volume={23},
  number={4},
  pages={3870-3885},
  doi={10.1109/TWC.2023.3312723}}

@book{sibille2010mimo,
  title={MIMO: from theory to implementation},
  author={Sibille, Alain and Oestges, Claude and Zanella, Alberto},
  year={2010},
  publisher={Academic Press}
}

@article{hashem2015critical,
  title={Critical and important factors related with enhancing wireless communication using mimo technology},
  author={Hashem, Soukaena H and Saleh, Hassan H},
  journal={Diyala Journal of Engineering Sciences},
  pages={42--63},
  year={2015}
}

@ARTICLE{6798744,
  author={Lu, Lu and Li, Geoffrey Ye and Swindlehurst, A. Lee and Ashikhmin, Alexei and Zhang, Rui},
  journal={IEEE Journal of Selected Topics in Signal Processing}, 
  title={An Overview of Massive MIMO: Benefits and Challenges}, 
  year={2014},
  volume={8},
  number={5},
  pages={742-758},
  doi={10.1109/JSTSP.2014.2317671}}

@ARTICLE{1341263,
  author={Sanayei, S. and Nosratinia, A.},
  journal={IEEE Communications Magazine}, 
  title={Antenna selection in MIMO systems}, 
  year={2004},
  volume={42},
  number={10},
  pages={68-73},
  doi={10.1109/MCOM.2004.1341263}}

@INPROCEEDINGS{7034415,
  author={Casu, George and Tută, Leontin and Nicolaescu, Ioan and Moraru, Cătălin},
  booktitle={2014 22nd Telecommunications Forum Telfor (TELFOR)}, 
  title={Some aspects about the advantages of using MIMO systems}, 
  year={2014},
  volume={},
  number={},
  pages={320-323},
  doi={10.1109/TELFOR.2014.7034415}}

@INPROCEEDINGS{1399142,
  author={Fishler, E. and Haimovich, A. and Blum, R. and Cimini, R. and Chizhik, D. and Valenzuela, R.},
  booktitle={Conference Record of the Thirty-Eighth Asilomar Conference on Signals, Systems and Computers, 2004.}, 
  title={Performance of MIMO radar systems: advantages of angular diversity}, 
  year={2004},
  volume={1},
  number={},
  pages={305-309 Vol.1},
  doi={10.1109/ACSSC.2004.1399142}}

@misc{xiong2023fundamentaltradeoffintegratedsensing,
      title={On the Fundamental Tradeoff of Integrated Sensing and Communications Under Gaussian Channels}, 
      author={Yifeng Xiong and Fan Liu and Yuanhao Cui and Weijie Yuan and Tony Xiao Han and Giuseppe Caire},
      year={2023},
      eprint={2204.06938},
      archivePrefix={arXiv},
      primaryClass={cs.IT},
      url={https://arxiv.org/abs/2204.06938}, 
}

@ARTICLE{10217169,
  author={Hua, Haocheng and Han, Tony Xiao and Xu, Jie},
  journal={IEEE Transactions on Wireless Communications}, 
  title={MIMO Integrated Sensing and Communication: CRB-Rate Tradeoff}, 
  year={2024},
  volume={23},
  number={4},
  pages={2839-2854},
  doi={10.1109/TWC.2023.3303326}}

@book{HeathJr._Lozano_2018, 
place={Cambridge},
title={Foundations of MIMO Communication},
publisher={Cambridge University Press},
author={Heath Jr., Robert W. and Lozano, Angel},
year={2018}}

@misc{liu2023deterministicrandomtradeoffintegratedsensing,
      title={Deterministic-Random Tradeoff of Integrated Sensing and Communications in Gaussian Channels: A Rate-Distortion Perspective}, 
      author={Fan Liu and Yifeng Xiong and Kai Wan and Tony Xiao Han and Giuseppe Caire},
      year={2023},
      eprint={2212.10897},
      archivePrefix={arXiv},
      primaryClass={cs.IT},
      url={https://arxiv.org/abs/2212.10897}, 
}

@ARTICLE{10129042,
  author={Ouyang, Chongjun and Liu, Yuanwei and Yang, Hongwen and Al-Dhahir, Naofal},
  journal={IEEE Communications Magazine}, 
  title={Integrated Sensing and Communications: A Mutual Information-Based Framework}, 
  year={2023},
  volume={61},
  number={5},
  pages={26-32},
  doi={10.1109/MCOM.001.2200493}}

@ARTICLE{10038611,
  author={Ouyang, Chongjun and Liu, Yuanwei and Yang, Hongwen},
  journal={IEEE Wireless Communications Letters}, 
  title={MIMO-ISAC: Performance Analysis and Rate Region Characterization}, 
  year={2023},
  volume={12},
  number={4},
  pages={669-673},
  doi={10.1109/LWC.2023.3238842}}

@ARTICLE{9800940,
  author={Ouyang, Chongjun and Liu, Yuanwei and Yang, Hongwen},
  journal={IEEE Wireless Communications Letters}, 
  title={Performance of Downlink and Uplink Integrated Sensing and Communications (ISAC) Systems}, 
  year={2022},
  volume={11},
  number={9},
  pages={1850-1854},
  doi={10.1109/LWC.2022.3184409}}

@ARTICLE{9705498,
  author={Liu, An and Huang, Zhe and Li, Min and Wan, Yubo and Li, Wenrui and Han, Tony Xiao and Liu, Chenchen and Du, Rui and Tan, Danny Kai Pin and Lu, Jianmin and Shen, Yuan and Colone, Fabiola and Chetty, Kevin},
  journal={IEEE Communications Surveys \& Tutorials}, 
  title={A Survey on Fundamental Limits of Integrated Sensing and Communication}, 
  year={2022},
  volume={24},
  number={2},
  pages={994-1034},
  doi={10.1109/COMST.2022.3149272}}

@ARTICLE{9606831,
  author={Cui, Yuanhao and Liu, Fan and Jing, Xiaojun and Mu, Junsheng},
  journal={IEEE Network}, 
  title={Integrating Sensing and Communications for Ubiquitous IoT: Applications, Trends, and Challenges}, 
  year={2021},
  volume={35},
  number={5},
  pages={158-167},
  doi={10.1109/MNET.010.2100152}}

@ARTICLE{10418473,
  author={Lu, Shihang and Liu, Fan and Li, Yunxin and Zhang, Kecheng and Huang, Hongjia and Zou, Jiaqi and Li, Xinyu and Dong, Yuxiang and Dong, Fuwang and Zhu, Jia and Xiong, Yifeng and Yuan, Weijie and Cui, Yuanhao and Hanzo, Lajos},
  journal={IEEE Internet of Things Journal}, 
  title={Integrated Sensing and Communications: Recent Advances and Ten Open Challenges}, 
  year={2024},
  volume={11},
  number={11},
  pages={19094-19120},
  doi={10.1109/JIOT.2024.3361173}}

@INBOOK{9963506,
  author={Zhang, J. Andrew and Guo, Yingjie Jay and Wu, Kai},
  booktitle={Joint Communications and Sensing: From Fundamentals to Advanced Techniques}, 
  title={Introduction to Joint Communications and Sensing (JCAS)}, 
  year={2023},
  volume={},
  number={},
  pages={1-30},
  doi={10.1002/9781119982944.ch1}}

@ARTICLE{9540344,
  author={Zhang, J. Andrew and Liu, Fan and Masouros, Christos and Heath, Robert W. and Feng, Zhiyong and Zheng, Le and Petropulu, Athina},
  journal={IEEE Journal of Selected Topics in Signal Processing}, 
  title={An Overview of Signal Processing Techniques for Joint Communication and Radar Sensing}, 
  year={2021},
  volume={15},
  number={6},
  pages={1295-1315},
  doi={10.1109/JSTSP.2021.3113120}}

@article{article,
author = {Cimini, D. and Marzano, Frank and Biscarini, Marianna and Gil, Rebeca and Schluessel, Peter and Concaro, Filippo and Marchetti, Matteo and Pasian, Marco and Romano, Filomena},
year = {2021},
month = {01},
pages = {1-1},
title = {Applicability of the Langley Method for Non-Geostationary In-Orbit Satellite Effective Isotropic Radiated Power Estimation},
volume = {PP},
journal = {IEEE Transactions on Antennas and Propagation},
doi = {10.1109/TAP.2020.3048479}
}

@INPROCEEDINGS{4411629,
  author={Ding, Minhua and Blostein, Steven D.},
  booktitle={IEEE GLOBECOM 2007 - IEEE Global Telecommunications Conference}, 
  title={Uplink-Downlink Duality in Normalized MSE or SINR Under Imperfect Channel Knowledge}, 
  year={2007},
  volume={},
  number={},
  pages={3786-3790},
  doi={10.1109/GLOCOM.2007.719}}

@misc{miretti2024uldldualitycellfreemassive,
      title={UL-DL duality for cell-free massive MIMO with per-AP power and information constraints}, 
      author={Lorenzo Miretti and Renato L. G. Cavalcante and Emil Björnson and Sławomir Stańczak},
      year={2024},
      eprint={2301.06520},
      archivePrefix={arXiv},
      primaryClass={cs.IT},
      url={https://arxiv.org/abs/2301.06520}, 
}

@INPROCEEDINGS{7510762,
  author={Zarei, Shahram and Gerstacker, Wolfgang and Schober, Robert},
  booktitle={2016 IEEE International Conference on Communications (ICC)}, 
  title={Uplink/downlink duality in massive MIMO systems with hardware impairments}, 
  year={2016},
  volume={},
  number={},
  pages={1-7},
  doi={10.1109/ICC.2016.7510762}}

@INPROCEEDINGS{5413251,
  author={Endeshaw, Tadilo and Chalise, Batu K. and Vandendorpe, Luc},
  booktitle={2009 3rd IEEE International Workshop on Computational Advances in Multi-Sensor Adaptive Processing (CAMSAP)}, 
  title={MSE uplink-downlink duality of MIMO systems under imperfect CSI}, 
  year={2009},
  volume={},
  number={},
  pages={384-387},
  doi={10.1109/CAMSAP.2009.5413251}}

@misc{attiah2024beamformingdesignintegratedsensing,
      title={Beamforming Design for Integrated Sensing and Communications Using Uplink-Downlink Duality}, 
      author={Kareem M. Attiah and Wei Yu},
      year={2024},
      eprint={2404.13392},
      archivePrefix={arXiv},
      primaryClass={cs.IT},
      url={https://arxiv.org/abs/2404.13392}, 
}

@INPROCEEDINGS{10615441,
  author={Perera, Thakshila and Mezghani, Amine and Hossain, Ekram},
  booktitle={2024 IEEE International Conference on Communications Workshops (ICC Workshops)}, 
  title={On the Characterization of Pareto Boundary for Joint Communication and Sensing in MIMO Wireless Systems}, 
  year={2024},
  volume={},
  number={},
  pages={685-690},
  doi={10.1109/ICCWorkshops59551.2024.10615441}}

@ARTICLE{6860253,
  author={Xu, Jie and Liu, Liang and Zhang, Rui},
  journal={IEEE Transactions on Signal Processing}, 
  title={Multiuser MISO Beamforming for Simultaneous Wireless Information and Power Transfer}, 
  year={2014},
  volume={62},
  number={18},
  pages={4798-4810}}
\end{document}